\documentclass[twocolumn]{aastex631}
\usepackage{amsmath}
\usepackage{float}
\usepackage{ulem}

\usepackage{CJK}
\newcommand{\ctext}[1]{\textup{\begin{CJK*}{UTF8}{bkai}#1\ignorespacesafterend\end{CJK*}}}

\newcommand{\alfven}{Alfv\'{e}n~}
\renewcommand{\vec}[1]{\boldsymbol{#1}}

\submitjournal{\apj}

\shorttitle{Inner PPD Turbulence}
\shortauthors{Rea et al.}

\begin{document}

\title{Magnetically Driven Turbulence in the Inner Regions of Protoplanetary Disks}

\author[0000-0002-5000-2747]{David G. Rea}
\affiliation{Department of Physics and Astronomy, Iowa State University, Ames, IA, 50010, USA}

\author[0000-0002-3771-8054]{Jacob B. Simon}
\affiliation{Department of Physics and Astronomy, Iowa State University, Ames, IA, 50010, USA}

\author[0000-0001-6259-3575]{Daniel Carrera}
\affiliation{Department of Physics and Astronomy, Iowa State University, Ames, IA, 50010, USA}

\author[0000-0002-8896-9435]{Geoffroy Lesur}
\affiliation{Univ. Grenoble Alpes, CNRS, IPAG, 38000 Grenoble, France}

\author[0000-0002-3768-7542]{Wladimir Lyra}
\affiliation{New Mexico State University, Department of Astronomy, PO Box 30001 MSC 4500, Las Cruces, NM 88003-8001, USA}

\author[0000-0003-0801-3159]{Debanjan Sengupta}
\affiliation{New Mexico State University, Department of Astronomy, PO Box 30001 MSC 4500, Las Cruces, NM 88003-8001, USA}

\author[0000-0003-2589-5034]{Chao-Chin Yang (\ctext{楊朝欽})}
\affiliation{Department of Physics and Astronomy, The University of Alabama,
    Box~870324, Tuscaloosa, AL~35487-0324, U.S.A.}

\author[0000-0002-3644-8726]{Andrew N. Youdin}
\affiliation{University of Arizona, Steward Observatory and LPL, Tucson, AZ 85719, USA}

\correspondingauthor{David G. Rea}
\email{drea1@iastate.edu}

%
%
\begin{abstract}

Given the important role turbulence plays in the settling and growth of dust grains in protoplanetary disks, it is crucial that we determine whether these disks are turbulent and to what extent. Protoplanetary disks are weakly ionized near the mid-plane, which has led to a paradigm in which largely laminar magnetic field structures prevail deeper in the disk, with angular momentum being transported via magnetically launched winds. Yet, there has been little exploration on the precise behavior of the gas within the bulk of the disk. We carry out 3D, local shearing box simulations that include all three low-ionization effects (Ohmic diffusion, ambipolar diffusion, and the Hall effect) to probe the nature of magnetically driven gas dynamics 1--30 AU from the central star. We find that gas turbulence can persist with a generous yet physically motivated ionization prescription (order unity Elsasser numbers). The gas velocity fluctuations range from 0.03--0.09 of the sound speed $c_s$ at the disk mid-plane to $\sim c_s$ near the disk surface, and are dependent on the initial magnetic field strength. However, the turbulent velocities do not appear to be strongly dependent on the field polarity, and thus appear to be insensitive to the Hall effect. The mid-plane turbulence has the potential to drive dust grains to collision velocities exceeding their fragmentation limit, and likely reduces the efficacy of particle clumping in the mid-plane, though it remains to be seen if this level of turbulence persists in disks with lower ionization levels.

\end{abstract}



%
%
\section{Introduction}
\label{sec:introduction}

Whether gas motions in protoplanetary disks are turbulent or laminar has significant implications for numerous steps in the planet formation process. For instance, if present, strong turbulence can cause large collisional velocities between dust grains, hindering their growth via fragmentation or bouncing \citep{Ormel2007,Guttler2010,Kothe2016}. Turbulence may concentrate particles in turbulent eddies and facilitate the first stages of planetesimal formation \citep{Cuzzi2008,Johansen2007,Hartlep2020} though it is also possible for turbulence to prevent or slow down the formation of planetesimals produced by the streaming instability (e.g. \citealt{Umurhan2020,Chen2020,Gole2020,Lim2023_arxiv}). Turbulence may also play an important role in the migration of protoplanets, by generating stochastic net torques that result in a random walk migration, as opposed to Type~I monotonic inward drift \citep{Nelson2004,Johnson2006,Yang2009,Paardekooper2011,Yang2012,Stoll2017}.

There are a number of mechanisms that may give rise to turbulence in disks, ranging from purely hydrodynamic (i.e., not requiring a magnetic field) mechanisms (see \citealt{Lyra2019}) to those induced by magnetic fields.  The most studied mechanism in the latter category is the magnetorotational instability (MRI; \citealt{Balbus1998}). The MRI arises naturally in sufficiently ionized Keplerian disks with a (relatively weak) magnetic field present and gives rise to correlated turbulent fluctuations in the gas velocity and magnetic field that produce an outward flux of angular momentum. However, most of the gas in protoplanetary disks is cold and poorly ionized, and as such, much of the disk is dominated by low-ionization physics (e.g., \citealt{Gammie1996,Wardle2007}): Ohmic diffusion (where electrons are collisionally impeded by neutral species, i.e. low gas conductivity), the Hall effect (current induced by the drift velocity between electrons and ions) and ambipolar diffusion (drift between the bulk neutral gas and ions). At a given magnetic field strength, gas becomes more decoupled from the field as density increases, i.e. increasing frequency of collisions between electrons, ions, and neutral species. In order of increasing gas density, the neutrals decouple (ambipolar diffusion), then ions (Hall effect), and then electrons (Ohmic dissipation). The importance of these effects have been studied extensively by both analytic \citep[e.g.,][]{Gammie1996,Balbus2001,Kunz2004,Desch2004,Salmeron2005} and numerical works \citep[e.g.,][]{Sano2002,Fleming2003,Lesur2014,Bai2014,Bai2015,Simon2015_numeric,Gressel2015,Bethune2017,Simon2018,Cui2021}.

Ohmic and ambipolar diffusion generally weaken MRI-driven turbulence. At large distances from the central star ($\sim 30$--100 AU, though the exact distance is very dependent on disk properties), a layered structure is thought to exist in which MRI-active surface regions ionized by far-ultraviolet (FUV) photons \citep{Perez-Becker2011} surround a mid-plane region in which ambipolar diffusion dampens MRI-driven turbulence \citep{Simon2013_strongacc,Simon2013_weakacc,Cui2021}. A similar structure has long been thought to be at work in the inner regions of disks but with Ohmic diffusion quenching the mid-plane region \citep{Gammie1996,Fleming2003}. This classic dead zone paradigm, which actually preceded the layered model for the outer disk, has been modified significantly over recent years with the appreciation that the Hall effect fundamentally alters the gas dynamics at intermediate radii (1--30 AU, \citealt{Wardle2007}).

When the Hall effect is important, the orientation of the vertical magnetic field plays a role in determining the ensuing gas dynamics \citep{Wardle1999,Balbus2001,Kunz2008,Bai2015,Simon2015_numeric}. In particular, the Hall-Shear Instability (HSI; \citealt{Kunz2008}) occurs when electromagnetic (whistler) waves couple to the background shear. The HSI is typically considered to be active only when the magnetic field points in the same direction as the gas rotation vector; more generally, it is strongest when the gas vorticity, wavevector, magnetic field share a mutual axis. Considering only the vertical field $\vec{B}$ and the background Keplerian shear $\vec{\Omega}$ (which typically shares an axis with the gas vorticity in disks), the coupling parameter can be expressed as $(\vec{\Omega}\cdot\vec{B})$, which we will use hereafter. The Hall effect can cause strong, persistent, and highly laminar magnetic torques which act on the gas to transport angular momentum \citep{Lesur2014,Bai2015,Simon2015_numeric} if the vertical component of the magnetic field is aligned with the angular momentum vector of the rotating disk gas $(\vec{\Omega}\cdot\vec{B} > 0)$, and non-axisymmetric modes of the HSI can produce intermittent bursts of magnetic torque \citep{Simon2015_numeric} even when the magnetic field is anti-aligned with the disk rotation $(\vec{\Omega}\cdot\vec{B} < 0)$, though these bursts can be quenched by stronger ambipolar diffusion \citep{Bai2015,Simon2015_numeric}. 

Another source of angular momentum transport in protoplanetary disks is magneto-thermal winds \citep{Bai2016_diskwind,Bai2017,Bethune2017}; in this process, gas is heated and ionized in the upper disk layers. This ionization couples the gas to the magnetic field, which then launches material away from the disk along magnetic field lines extending beyond the disk surface, transporting angular momentum in the process as in the Blandford-Payne mechanism \citep{Blandford1982}. The importance of this mechanism in angular momentum transport, the critical role that non-ideal MHD effects play on regulating magnetic activity, and the (resulting) largely laminar magnetic fields within the disk \cite[e.g.,][]{Lesur2014} has led to a new paradigm where accretion is assumed to be a largely laminar process, in contrast to the vigorous turbulence induced by the MRI.

Despite the dominance of largely laminar magnetic stresses in angular momentum transport, it is still not well understood how exactly the gas behaves in the inner disk ($<30$ AU) where the Hall effect is dominant.  Indeed, as shown by \cite{Simon2018}, at $\sim 30$--100AU scales, the gas can be rendered turbulent for a large range of parameter values, even if magneto-thermal winds dominate the angular momentum transport. This was found to be caused by zonal flows that create regions both stable and unstable to the MRI; the latter regions generate turbulece that then propagates throughout the domain  (see also \citealt{Cui2021}, which confirm the presence of outer disk turbulence with global simulations).

Here we ask whether similar processes can be at work in the inner disk. While the results of \cite{Bai2015} suggest (see their Fig. 6 for instance) that there may be some form of gas turbulence at these smaller radii, there are three questions which still need to be addressed in order to elucidate the nature of gas motions in the inner disk:

\begin{enumerate}
    \item What is the strength of non-laminar gas motions (if present) in the 1--30AU regions of protoplanetary disks?
    \item Do these gas motions resemble disk turbulence as seen in earlier simulations such as in \citealt{Simon2012} or \citealt{Simon2015_alma})?
    \item What is the origin of these gas motions?
\end{enumerate}

In this paper, we will address the first two questions, with a focus on characterizing gas motions as a function of the relevant parameters (e.g., magnetic field strength, orientation). We carry out a series of 3D, local shearing box simulations that employ fully non-ideal MHD; while such a setup cannot accurately capture magneto-thermal winds (e.g., \citealt{Fromang2013}), our focus is on the gas motions themselves, particularly near the mid-plane where planet formation occurs.

This paper is organized as follows: in \S\ref{sec:method}, we describe our simulations and methods in detail. In \S\ref{sec:results}, we present the magnetic field and velocity structure from our fiducial simulation.  In \S\ref{sec:discussion}, we examine the remainder of our parameter space and investigate the possible sources of turbulent gas motions. In \S\ref{sec:conclusions}, we summarize our conclusions and comment on their implications for protoplanetary disk evolution and planetesimal formation.

%
%
\section{Method}
\label{sec:method}

\subsection{Shearing Box}
\label{sec:method:shearing_box}

We carry out our numerical simulations using Athena, a second-order accurate dimensionally-unsplit Godunov code for solving the equations of MHD \citep{Stone2008}. We use the HLLD Reimann solver of \citet{Miyoshi2005} with the piecewise-parabolic method \citep[PPM;][]{Colella1984} for third order spatial reconstruction and the corner-transport-upwind method \citep[CTU;][]{Colella1990} time integrator. The constraint $\boldsymbol\nabla\cdot\mathbf{B} = 0$ is maintained to machine precision via the constrained transport algorithm \citep[CT;][]{Evans1988}. Non-ideal MHD terms are implemented using the methods of \citet{Simon2009_resistivity} for Ohmic resistivity, \citet{Bai2011_ambipolar} for ambipolar diffusion, and \citet{Bai2014} for the Hall effect.

We use the local shearing box approximation  (e.g., \citealt{Hawley1995,Stone2010}) for our computational domain in order to resolve turbulence on scales $\ll H$ (where $H$ is the gas scale height) with as many grid cells as possible with our current computational resources. The shearing box is a small patch of the disk at a fiducial radius $R_0$ and co-orbiting the central star with Keplerian angular velocity $\mathbf\Omega = \Omega_0\hat{\vec{e}}_z$. The shearing box has Cartesian coordinates $(x, y, z) = (R-R_0$, $R_0\phi$, $z)$, where $R$ and $\phi$ correspond to the radial and azimuthal directions of a cylindrical coordinate system, respectively. To account for differential rotation, there is a linear background shear $\vec{v}_0 = -q\Omega_0x\hat{\vec{e}}_y$, where $q$ is the shear parameter:

\begin{align}
    q &= -\left. \frac{d\ln{\Omega}}{d\ln{R}} \right|_{R=R_0}.
\end{align}

\noindent
The shear parameter $q=3/2$ for a Keplerian disk. 

The shearing box boundaries are periodic in $y$, shear-periodic in $x$ \citep{Hawley1995,Stone2010}, and a modified outflow boundary in $z$ \citep{Simon2011_stratified}. Shearing box simulations of the MRI in vertically stratified disks exhibit significant mass outflow through the vertical domain boundaries \citep{Suzuki2009,Simon2013_strongacc}; we continuously replenish disk mass to compensate for mass loss as described in \citet{Simon2015_numeric}. We also employ the Crank-Nicolson method \citep{Stone2010} for the non-inertial terms associated with the shearing box setup.  Finally, we employ the orbital advection algorithm, as described in \cite{Stone2010} to subtract off the Keplerian shear and integrate this component analytically. Thus, in the equations below, the velocity field is defined to have this shear already subtracted.

In units such that the magnetic permeability is unity, the equations of MHD are the continuity equation

\begin{equation} \label{eq:continuity}
    \frac{\partial\rho}{\partial t} + \vec{\nabla}\cdot(\rho \vec{v}) = 0,
\end{equation}

\noindent
the momentum equation

\begin{multline} \label{eq:momentum}
    \frac{\partial\rho \vec{v}}{\partial t} + \vec{\nabla}\cdot(\rho\vec{vv} - \vec{BB}) + \vec{\nabla}\left(P + \frac{1}{2}B^2\right) \\
     = 2\rho q\Omega_0^2\vec{x} - \rho\Omega_0^2\vec{z} - 2\Omega_0\vec{\hat{e}}_z\times\rho\vec{v},
\end{multline}

\noindent
and the induction equation

\begin{multline} \label{eq:induction}
    \frac{\partial\vec{B}}{\partial t} - \vec\nabla\times(\vec{v}\times \vec{B}) \\
    = -\vec\nabla\times\left[ \eta_O\vec{J} + \eta_H\frac{\vec{J}\times \vec{B}}{B} - \eta_A\frac{(\vec{J}\times \vec{B})\times \vec{B}}{B^2} \right].
\end{multline}

\noindent
Here $\rho$ is the mass density, $\vec{v}$ is the gas velocity, $\vec{B}$ is the magnetic field, and $\vec{J} = \vec\nabla\times \vec{B}$ is the current density. In this work, the magnetic field is normalized by a factor $\sqrt{4\pi}$. From left to right, the R.H.S. terms of the momentum equation represent radial tidal forces, vertical gravity, and the Coriolis force. We assume an isothermal equation of state with gas pressure $P=\rho c_s^2$, where $c_s$ is the isothermal sound speed $\sqrt{k_BT/\mu m_p}$, with Boltzmann constant $k_B$, disk temperature $T$, proton mass $m_p$ and mean molecular weight $\mu = 2.33$ (see Section~\ref{sec:method:diskmodel} for specific values of these quantities). The parameters $\eta_O$, $\eta_H$, and $\eta_A$ are the Ohmic, Hall, and ambipolar diffusion coefficients, respectively. When the only charged species are electrons and ions (i.e. in the absence of charged dust grains) and ignoring the effect of disk chemistry on ionization, these are \citep{Balbus2001,Wardle2007}:

\begin{align}
    \eta_O &= \frac{c^2 m_e}{4\pi e^2}\frac{n}{n_e}\langle\sigma v\rangle_e \label{eq:eta_Om} \\
    \eta_H &= \frac{Bc}{\sqrt{4\pi}en_e} \label{eq:eta_Ha} \\
    \eta_A &= \frac{B^2}{\gamma_i\rho\rho_i} \label{eq:eta_AD}
\end{align}

\noindent
where $m_e$ is the electron mass, $n$ and $n_e$ are the number densities of the neutrals and the electrons, respectively, and $\rho_i$ is the ion mass density. 
\begin{equation}
    \langle\sigma v\rangle_e = 8.28\times 10^{-9}\left(\frac{T}{\text{100 K}}\right)^{1/2} \text{ cm$^3$ s$^{-1}$}
\end{equation}

\noindent
is the electron-neutral collision rate \citep{Draine1983}, $\gamma_i = \langle\sigma v\rangle_i/(m_n + m_i)$, and

\begin{equation}
    \langle\sigma v\rangle_i = 1.3\times 10^{-9} \text{ cm$^3$ s$^{-1}$}
\end{equation}

\noindent
is the ion-neutral collision rate \citep{Draine2011}. Here, $m_i$ and $m_n$ are the masses of the ions and neutrals, respectively.

\subsection{Disk Model}
\label{sec:method:diskmodel}

\begin{deluxetable}{ccc}
    \tablecaption{Disk Model Parameters \label{table:disk_model}}
    \tablewidth{0pt}
    \tablehead{\colhead{Parameter } & \colhead{ $\left.\text{\hspace{100pt}}\right.$} &\colhead{Value}}
    \startdata
        $M_{\rm star}$ & & $1.0 M_\odot$ \\
        $M_{\rm disk}$ & & $0.05 M_\odot$ \\
        $R_c$ & & $100$ AU \\
        $\gamma$ & & 1.0 \\
        $T_X$ & & 3.0 keV \\
        $L_X$ & &  $10^{30}$ erg s$^{-1}$ \\
        $\Sigma_{\rm FUV}$ & & $0.005$ g cm$^{-2}$ \\
    \enddata
\end{deluxetable}

We center our local simulations at specific radii in a model disk, which is described as follows.
We consider an exponentially tapered surface density profile as in \cite{Lynden-Bell1974} and \cite{Hartmann1998},

\begin{multline}
    \Sigma(R) = \frac{M_{\rm disk}(2-\gamma)}{2\pi R_c^2} \\
    \times \left( \frac{R}{R_c} \right)^{-\gamma} \exp\left[ -\left( \frac{R}{R_c} \right)^{2-\gamma} \right] \text{ g cm$^{-2}$},
\end{multline}

\noindent
where $M_{\rm disk}$ is the disk mass, $R_c$ is the characteristic radius of the disk, and $\gamma$ is the surface density power law index. The values of these model parameters are chosen to be representative of the disks observed by \cite{Andrews2009}, and can be found in Table \ref{table:disk_model}. The temperature profile is

\begin{equation}
    T(R) = 280 \left( \frac{R}{1 \text{ AU}} \right)^{-1/2} \text{ K}.
\end{equation}

With this temperature structure, vertical hydrostatic equilibrium requires a density profile

\begin{equation}
    \rho(x, y, z) = \rho(R_0) \exp\left( -\frac{z^2}{2H^2} \right)
\end{equation}

\noindent
with scale height $H = c_s/\Omega_0$.

\subsection{Ionization}
\label{sec:method:ionization}

We include contributions to the ionization rate from radioactive decay with $\zeta_{\rm RD} = 10^{-19}$ s$^{-1}$ \citep{Umebayashi2009} and cosmic rays with $\zeta_{\rm CR} = \zeta_{\rm CR,0}\exp(-\Sigma/96 \text{ g cm}^{-2})$ s$^{-1}$ \citep{Umebayashi1981}. The cosmic ray flux ($ \zeta_{\rm CR,0}$) is highly uncertain, but observations in the direction of $\zeta$ Persei suggest $\zeta_{\rm CR,0} \sim 10^{-16}$ \citep{McCall2003}, which we take to be our fiducial value in order to make contact with previous work (e.g. \citealt{Lesur2014}), though we note that some modeling predicts $\zeta_{\rm CR,0} \lesssim 10^{-19}$ in TW Hya \citep{Cleeves2015}. We include the contribution from stellar X-rays with $L_X = 10^{30}$ erg s$^{-1}$ and temperature $T_X = 3.0$ keV  \citep{Igea1999,Bai2009} as:

\begin{align}
    \zeta_{\rm XR} &= \left(\frac{L_X}{10^{29}\rm \;erg\;s^{-1}}\right) \left(\frac{R}{\rm 1\;AU}\right)^{-2.2} \nonumber\\
    &\times \left( \zeta_1\left[e^{-(N_{\rm H1}/N_1)^a}+e^{-(N_{\rm H2}/N_1)^a}\right] \right. \nonumber\\
    &\;\;+ \left. \zeta_2\left[e^{-(N_{\rm H1}/N_2)^b}+e^{-(N_{\rm H2}/N_2)^b}\right] \right)
\end{align}

\noindent
where $N_{\rm H1,2}$ is the column density of hydrogen integrating through the disk from above ($N_{\rm H1}$) and from below ($N_{\rm H2}$), and for $T_X = 3.0$ keV, $\zeta_1 = 6\times 10^{-12}{\rm\;s^{-1}}$, $N_1 = 1.5\times 10^{21}{\rm\;cm^{-2}}$, $a = 0.4$, $\zeta_2 = 1.0\times 10^{-15}{\;\rm s^{-1}}$, $N_2 = 7.0\times 10^{23}{\rm\;cm^{-2}}$, and $b = 0.65$. The X-ray ionization rate is only weakly sensitive to $T_X$ \citep{Igea1999}. 

We follow \citet{Bai2014} and treat FUV ionization as another independent ionization source with

\begin{equation}
    \zeta_{\rm FUV} = 10^{-6}\left(\frac{R}{1\;AU}\right)^{-2}\exp\left(-\frac{\Sigma}{\Sigma_{\rm FUV}}\right) \text{s}^{-1}
\end{equation}

\noindent
and $\Sigma_{\rm FUV} = 0.005$ g cm$^{-2}$. The ionization fraction $x_e$ is obtained by balancing these ionization sources with dissociative recombination in the absence of metals and dust \citep{Gammie1996,Fromang2002}:

\begin{equation} \label{eq:ionization}
    x_e = g\frac{n_e}{n_n} = g\sqrt{\frac{\zeta_{\rm RD} + \zeta_{\rm CR} + \zeta_{\rm XR} + \zeta_{\rm FUV}}{n_n\alpha_{dr}}},
\end{equation}

\noindent
where $\alpha_{dr} = 3\times 10^{-6}~(T/{\rm K})^{-1/2}$ cm$^2$ s$^{-1}$ is the dissociative recombination rate coefficient. Similar to \citet{Bai2017}, we multiply the ionization fraction by a term $g$, where

\begin{align}
    g = \exp\left[ \frac{0.3\Sigma_{\rm FUV}}{\Sigma + 0.01\Sigma_{\rm FUV}} \right].
\end{align}

\begin{figure}
    \centering
    \includegraphics[width=\columnwidth]{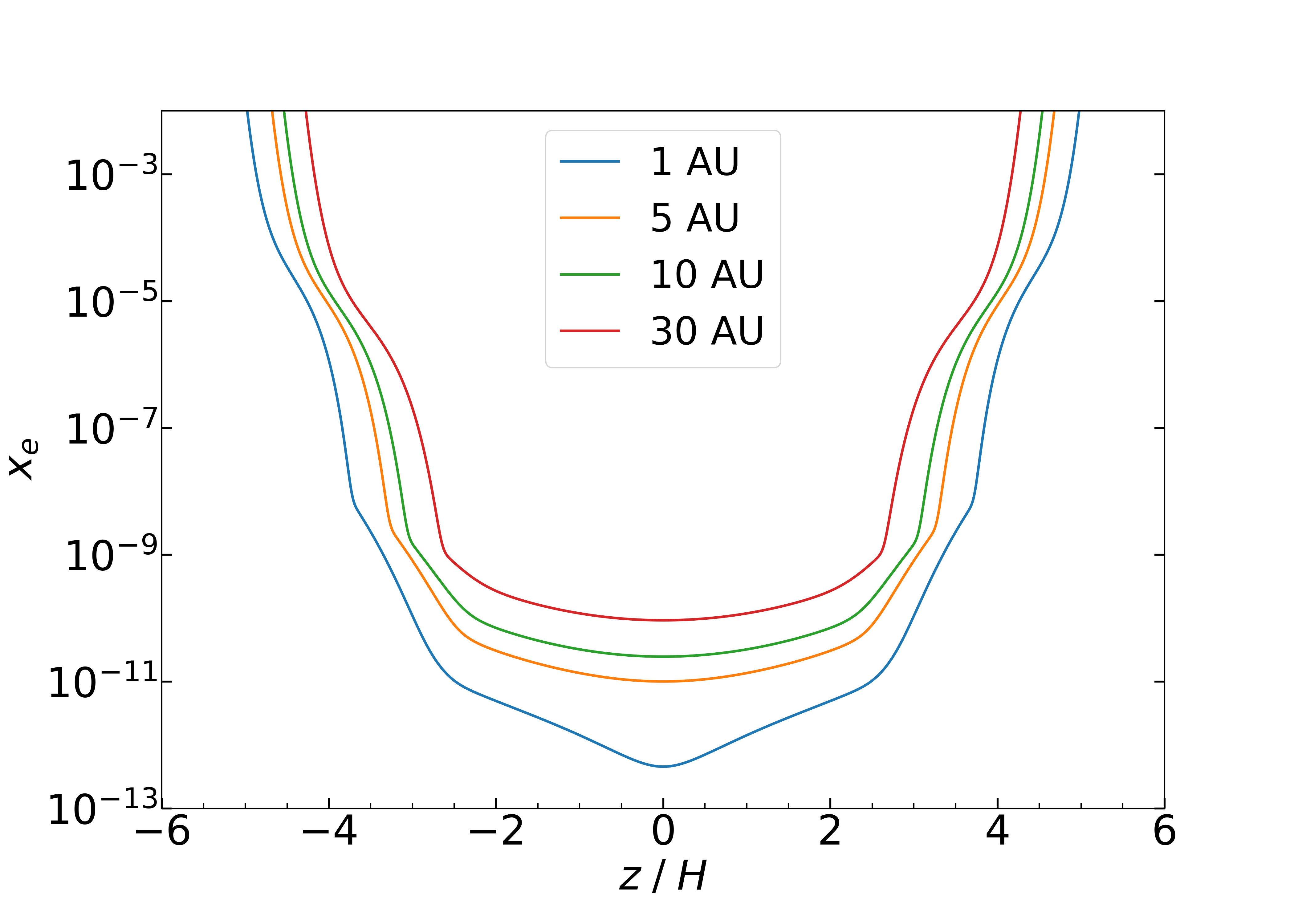}
    \caption{Initial ionization fraction $x_e$ as a function of height. The ionization near the disk surface ($|z|\gtrsim 3$ H) is dominated by FUV photons and by cosmic rays near the disk mid-plane.}
    \label{fig:ionization_fraction}
\end{figure}

\noindent
Without this term, ambipolar diffusion becomes increasingly important in the FUV-ionized layer, but this is not the case as shown by more sophisticated photochemistry calculations \citep{Walsh2012}. The resulting ionization fraction is shown in Figure \ref{fig:ionization_fraction} for various radii.

We quantify the diffusivity coefficients $\eta_O$, $\eta_H$, and $\eta_A$ (Equations \ref{eq:eta_Om}--\ref{eq:eta_AD}) via the dimensionless Elsasser numbers

\begin{align}
    \Lambda &= \frac{v_A^2}{\Omega_0\eta_O}, \\
    \text{Ha} &= \frac{v_A^2}{\Omega_0\eta_H}, \\
    \text{Am} &= \frac{v_A^2}{\Omega_0\eta_A},
\end{align}

\noindent
where $v_A$ is the magnitude of the \alfven velocity

\begin{equation}
    \boldsymbol{v_A} = \frac{\boldsymbol{B}}{\sqrt{\rho}},
\end{equation}

\begin{figure}
    \centering
    \includegraphics[width=\columnwidth]{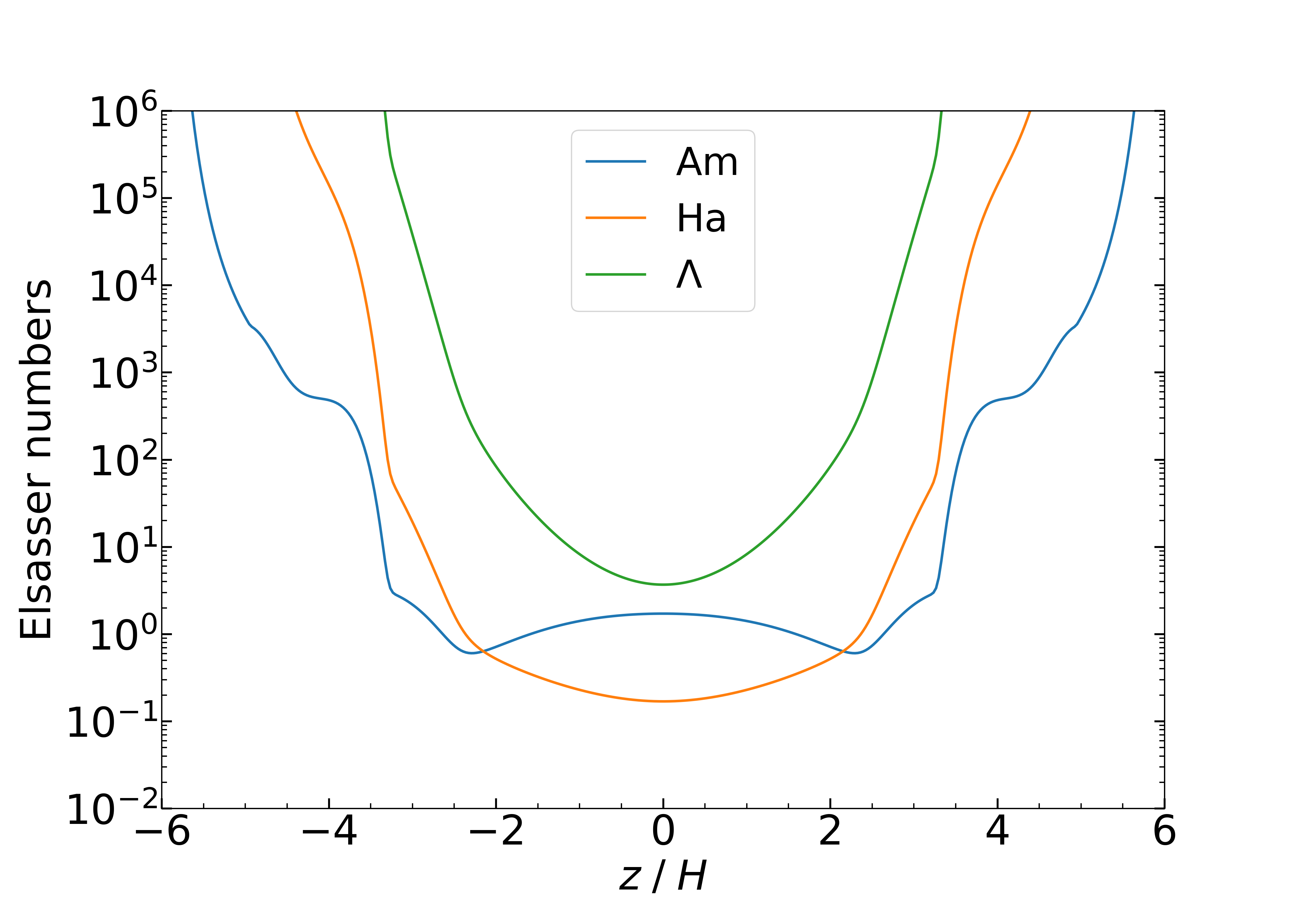}
    \caption{Initial Ohmic ($\Lambda$), Hall (Ha), and ambipolar (Am) Elsasser numbers as a function of height, when radius $R~=~5$~AU and $\beta_0~=~10^4$. A smaller Elsasser number indicates a stronger non-ideal MHD effect. Here, the Hall effect is the strongest non-ideal effect within $|z| < 2H$.}
    \label{fig:elsasser_profiles}
\end{figure}

\noindent
The initial vertical profiles of $\Lambda$, Ha, and Am are shown in Figure \ref{fig:elsasser_profiles}.

The disk ionization prescription adopted by this work is generous. Because the cosmic ray flux through protoplanetary disks is highly uncertain (disks may be shielded from cosmic ray flux by, e.g., stellar winds \citealt{Cleeves2013,Cleeves2015}), we also examine a setup with cosmic ray flux weaker by a factor of 10 (R5--b4p--weak-cr, see Table \ref{table:simulations}) as in \citet{Umebayashi1981}.

\subsection{Set of Simulations}
\label{sec:method:setup}

We assume the shearing box domain is threaded by a vertical magnetic field $\mathbf{B} = B_0\hat{\mathbf{e}}_z$ whose initial strength is specified by the signed ratio between the gas and magnetic pressures

\begin{equation} \label{eq:beta}
    \beta_0 \equiv \frac{B_0}{|B_0|}\frac{2\rho_0 c_{s,0}^2}{B_0^2},
\end{equation}

\noindent
where $\rho_0$ is the initial mid-plane density of the disk (and we define $P_0 = \rho_0c_s^2$ to be the initial mid-plane pressure). A positive $\beta_0$ $(+)$ means the initial vertical magnetic field is aligned with the disk angular momentum vector ($\vec{\Omega}\cdot\vec{B_0} > 0$), and a negative $\beta_0$ $(-)$ refers to the anti-aligned case ($\vec{\Omega}\cdot\vec{B_0} < 0$).

We explore radii of 1, 5, 10, and 30 AU. We take a fiducial $\beta_0 = +10^4$ (which corresponds to $|B_{z,0}| = 0.1$, 0.01, 0.005, and 0.001 G at 1, 5, 10, and 30 AU, respectively) and explore a strong ($\beta_0 = +10^3$, $|B_{z,0}| = 0.04$ G) and weak ($\beta_0 = +10^5$, $|B_{z,0}| = 0.004$ G) magnetic field at 5 AU. We also explore anti-aligned field orientations where $\beta_0 < 0$ at 5, 10, and 30 AU, for a field strength $|\beta_0|~=~10^4$. We set $c_s~=~\Omega_0 = \rho_0 = 1$ in dimensionless code units. 

The dimensions for each simulation are ($L_x, L_y, L_z$) = ($4H, 8H, 12H$) with a resolution of 32 grid zones per scale height $H$. This resolution is comparable to but slightly higher than previous 3D shearing box simulations that include the Hall effect alongside Ohmic and ambipolar diffusion (e.g., \citealt{Lesur2014,Bai2015,Simon2015_numeric,Simon2018}). The full set of simulations can be found in Table \ref{table:simulations}.

%
%
\section{Results}
\label{sec:results}

\begin{deluxetable*}{lccccccccc}
    \tablecaption{Shearing Box Simulations \label{table:simulations}}
    \colnumbers
    \tablehead{
        \colhead{Label} &
        \colhead{$R_{\rm AU}$} &
        \colhead{$\beta_0$} &
        \colhead{$\log_{10}\zeta_{\rm CR,0}$} &
        \colhead{${\rm Am_0}$} &
        \colhead{${\rm Ha_0}$} &
        \colhead{${\rm \Lambda_0}$} &
        \colhead{$\overline{\alpha}$} & 
        \colhead{$\overline{\alpha}_{|z|<H}$} &
        \colhead{$\overline{\langle\delta v/c_s\rangle}_{|z|<H}$}
    }
    \startdata
        R1--b4p & 1 & $+10^4$ & $-16$ & 0.27 & 0.0065 & 0.023 & $2.26\;(-3)$ & $7.86\;(-3)$ & $7.50\;(-2)$ \\
        R5--b3p & 5 & $+10^3$ & $-16$ & 1.7 & 0.54 & 37 & $4.91\;(-3)$ & $6.43\;(-3)$ & $5.79\;(-2)$ \\
        R5--b4p & 5 & $+10^4$ & $-16$ & 1.7 & 0.17 & 3.7 & $1.16\;(-3)$ & $1.91\;(-3)$ & $6.80\;(-2)$ \\
        R5--b4n & 5 & $-10^4$ & $-16$ & 1.7 & 0.17 & 3.7 & $7.95\;(-4)$ & $5.21\;(-4)$ & $5.21\;(-2)$ \\
        R5--b4p--weak-cr & 5 & $+10^4$ & $-17$ & 0.55 & 0.054 & 1.2 & $1.19\;(-3)$ & $1.57\;(-3)$ & $5.15\;(-2)$ \\
        R5--b5p & 5 & $+10^5$ & $-16$ & 1.7 & 0.054 & 0.37 & $3.13\;(-4)$ & $6.09\;(-4)$ & $3.27\;(-2)$ \\
        R10--b4p & 10 & $+10^4$ & $-16$ & 2.4 & 0.44 & 22 & $1.34\;(-3)$ & $2.06\;(-3)$ & $5.80\;(-2)$ \\
        R10--b4n & 10 & $-10^4$ & $-16$ & 2.4 & 0.44 & 22 & $1.06\;(-3)$ & $9.22\;(-4)$ & $6.28\;(-2)$ \\
        R30--b4p & 30 & $+10^4$ & $-16$ & 3.2 & 1.7 & 320 & $1.82\;(-3)$ & $2.37\;(-3)$ & $3.81\;(-2)$ \\
        R30--b4n & 30 & $-10^4$ & $-16$ & 3.2 & 1.7 & 320 & $1.69\;(-3)$ & $2.17\;(-3)$ & $8.84\;(-2)$ \\
    \enddata
    \tablecomments{Table of parameters and key diagnostics for the shearing box simulations: (1) Simulation label; (2) Radial location of the shearing box in AU; (3) Initial vertical magnetic field strength, see Equation \ref{eq:beta}; (4) Cosmic ray ionization rate; (5-7) Initial mid-plane values of the ambipolar, Hall, and Ohmic Elsasser numbers, respectively; (8) time-averaged $\alpha$ parameter, see Equation \ref{eq:alpha}; (9) Same as column 7, but the volume average is computed within $z = \pm {\rm H}$; (10) time- and volume-averaged turbulent velocity as a fraction of the sound speed, computed within $z = \pm {\rm H}$. Values written as $a\;(b)$ denote $a\times 10^b$.}
\end{deluxetable*}

We have run local simulations centered at several radii between 1 and 10 AU in our model disk; in this region, the Hall effect is the dominant non-ideal process at the mid-plane \citep[or see Figure \ref{fig:elsasser_profiles} in this work]{Kunz2004,Wardle2007}. We also run two simulations at 30 AU where the Hall effect is less dominant but still present \citep{Simon2015_numeric,Bai2015}.

\subsection{Angular Momentum Transport}
\label{sec:resutls:am_transport}

\begin{figure}
    \centering
    \includegraphics[width=\columnwidth]{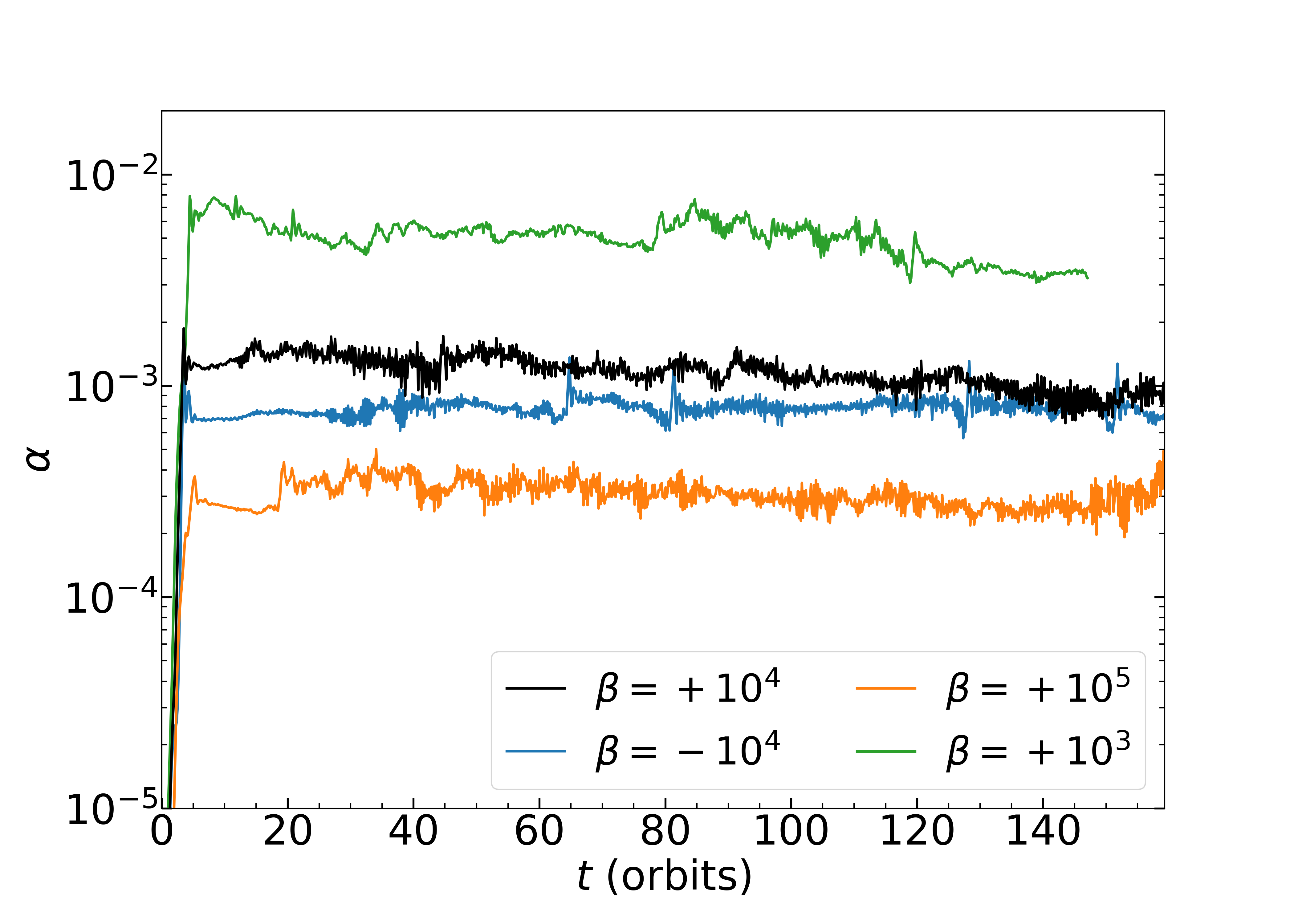}
    \caption{\citet{Shakura1973} $\alpha$ parameter over time for different field strengths and orientations (aligned with the disk rotation, $+$, and anti-aligned, $-$) at 5 AU. The green, black, blue, and orange curves correspond to $\beta_0 = +10^3, +10^4, -10^4, +10^5$ respectively. In all cases, the $\alpha$ stress exhibits small-amplitude fluctuations over time. Furthermore, $\alpha$ decreases with increasing $|\beta_0|$ and is also lower for the anti-aligned run compared with the aligned field run of the same field strength.}
    \label{fig:alpha}
\end{figure}

We examine the $R\phi$-component of the stress tensor, which is responsible for radial angular momentum transport. We quantify this in terms of the \citet{Shakura1973} $\alpha$ parameter.

\begin{align} \label{eq:alpha}
    \alpha = \frac{\langle\rho v_xv_y - B_xB_y\rangle}{\langle\rho c_s^2\rangle},
\end{align}

\noindent
where angled brackets denote the volume average

\begin{align}
    \langle Q\rangle = \frac{1}{L_xL_yL_z}\int Q(x,y,z)\;dxdydz
\end{align}

\noindent
When $|\beta_0| = 10^4$, we find $\overline{\alpha} \sim 10^{-3}$ (Figure \ref{fig:alpha}; Table \ref{table:simulations}), where the overline denotes the time average 

\begin{align}
    \overline{Q} = \frac{1}{\tau}\int Q(t)\;dt
\end{align}

\noindent
We average over times when $\alpha$ is in a statistical steady state, which corresponds to the last $\sim~140$ orbits for all but the run at 1 AU; in that run, we average over the last $\sim~35$ orbits. Our simulations with a stronger (weaker) initial magnetic field exhibit distinctly larger (smaller) $\alpha$. The $\alpha$ stresses exhibit small-amplitude fluctuations over time, typically by a factor of $\lesssim 2$, which may indicate some level of turbulent angular momentum transport in our simulations.

We compute the approximate corresponding mass accretion rate ignoring any contribution from disk winds and assuming a steady-state disk:

\begin{align}
    \dot{M} = 3\pi\alpha c_s H \Sigma
\end{align}

\noindent
At 5 AU, our disk model suggests that an $\alpha = 10^{-3}$ drives a mass accretion rate of $\dot{M} \approx 5\times 10^{-9}\;{\rm M_\odot/yr}$.  This value should be treated as a lower limit because we ignore the contribution from magnetic winds.

\subsection{Magnetic Field Structure and Evolution}
\label{sec:results:magnetic_field_structure}

\begin{figure*}
    \centering
    \includegraphics[width=\linewidth]{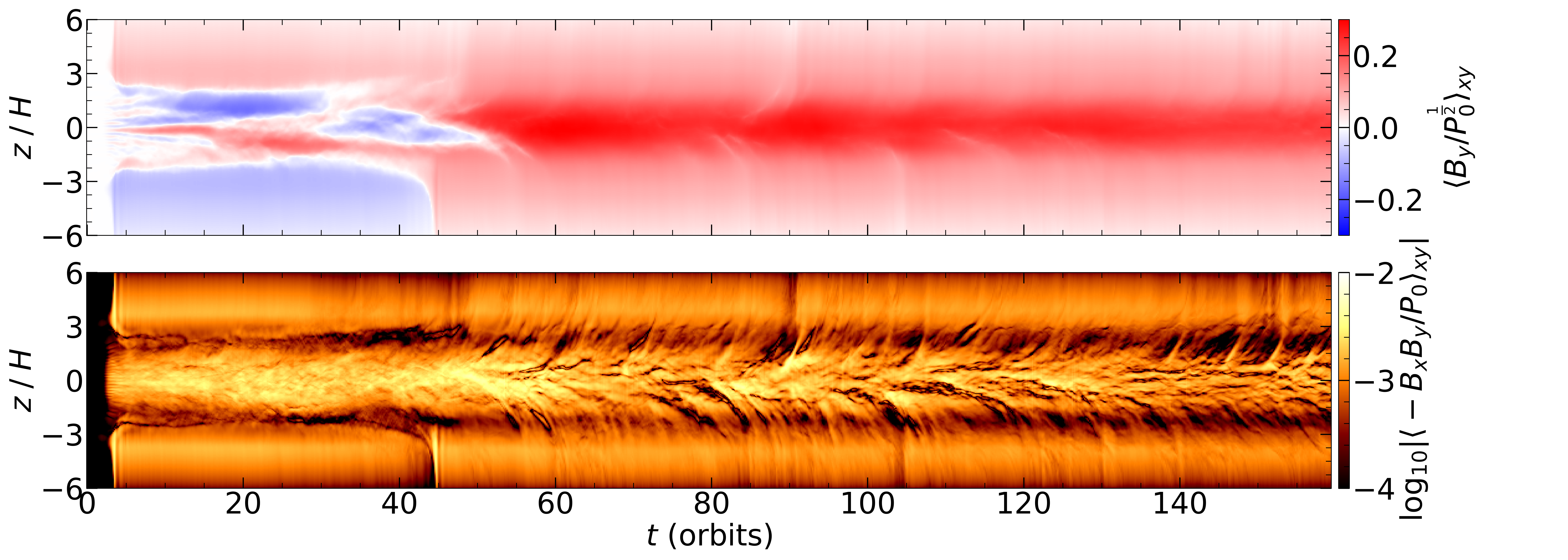}
    \caption{Spacetime diagram of $\langle B_y/\sqrt{P_0}\rangle_{xy}$ (top) and $\log_{10}\langle-B_xB_y/P_0\rangle_{xy}$ (bottom) for the run with $R = 5$ AU and $\beta_0 = 10^4$. The toroidal field is characterized by an even symmetry ($B_y(z) \simeq B_y(-z)$) after 50 orbits. The Maxwell stress is large throughout the vertical extent of the disk. The disk surfaces ($|z| \gtrsim 3$ H) exhibit a large-scale, mostly laminar Maxwell stress; this is distinct from the mid-plane region of the disk, which exhibits Maxwell stress with small-scale structure.}
    \label{fig:st_5AU_beta4}
\end{figure*}

\begin{figure*}
    \centering
    \includegraphics[width=\textwidth]{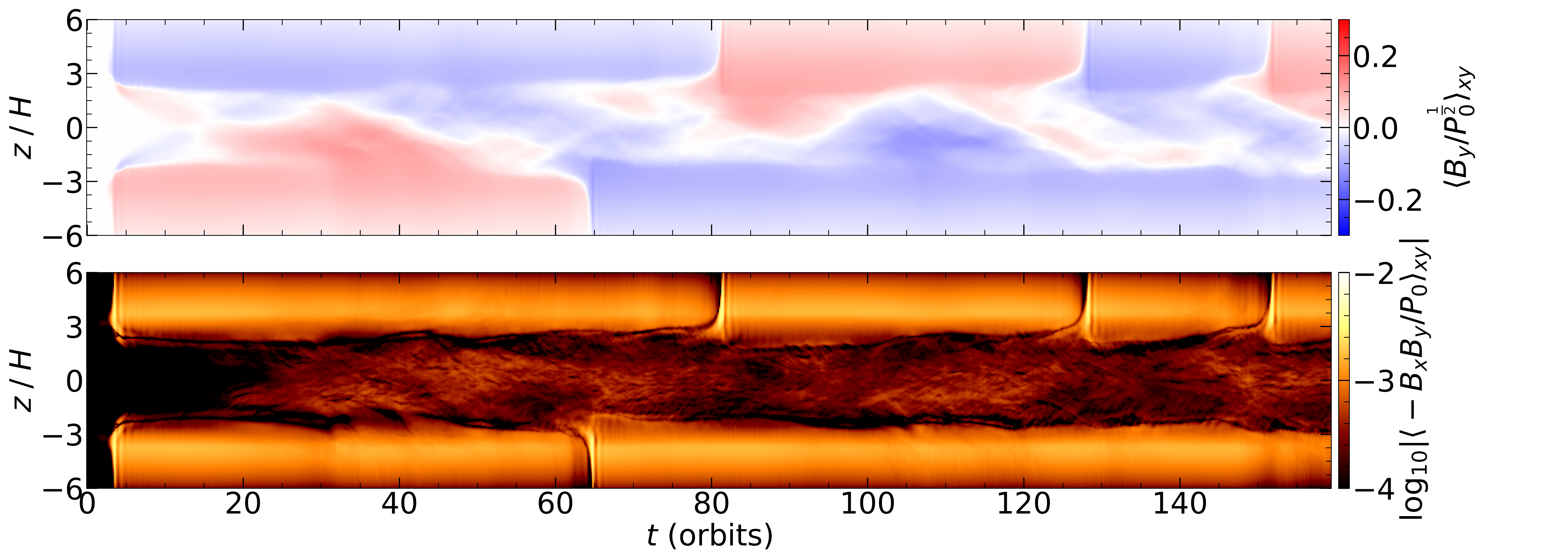}
    \caption{Spacetime diagram of $\langle B_y/\sqrt{P_0}\rangle_{xy}$ (top) and $\log_{10}\langle-B_xB_y/P_0\rangle_{xy}$ (bottom) for the run with $R = 5$ AU and $\beta_0 = -10^4$. The toroidal field changes sign across an approximately horizontal ($xy$-plane) current sheet. Similar to the ($\beta_0 > 0$) case, the Maxwell stress is large near the disk surface; however, in the mid-plane the Maxwell stress is largely suppressed.}
    \label{fig:st_5AU_beta4n}
\end{figure*}

We present the spacetime diagram of the toroidal (azimuthal) magnetic field $\langle B_y\rangle_{xy}$ and the corresponding Maxwell stress $\langle-B_xB_y\rangle_{xy}$, where $\langle\cdot\rangle_{xy}$ denotes the horizontal average;

\begin{align}
    \langle Q\rangle_{xy} = \frac{1}{L_xL_y}\int Q(x,y,z)\;dxdy
\end{align}

\noindent
for the runs with $R = 5$ AU and $\beta_0 = 10^4$ (Figure \ref{fig:st_5AU_beta4}) and $R = 5$ AU and $\beta_0 = -10^4$ (Figure \ref{fig:st_5AU_beta4n}).
In both the aligned and anti-aligned cases there is strong, large-scale Maxwell stress close to the disk surface. In the mid-plane of the disk, the Maxwell stress has small-scale vertical fluctuations in intensity. Additionally, the mid-plane Maxwell stress is extremely sensitive to the magnetic field polarity. In the aligned case, the Maxwell stress remains strong in both the mid-plane and surface regions of the disk, but in the anti-aligned case the mid-plane Maxwell stress ($|z| \lesssim 2$ H) is suppressed by a factor $\sim 4$.

The qualitative behavior of the toroidal field $\langle B_y\rangle_{xy}$ in our simulations also changes significantly between aligned and anti-aligned $B_{z,0}$. Both simulations exhibit thin sheets of current generated between regions of magnetic field with opposing sign, but here the similarities end. In the aligned case, multiple current sheets are present at the beginning of the simulation, though all are transient and last for fewer than $\sim 50$ orbits before the toroidal field organizes into an ``even" configuration [$B_y(z) \sim B_y(-z)$]. In the anti-aligned case, there is typically only one current sheet present at any time, and it does not remain at the mid-plane at all times. The toroidal field is organized into an ``odd" configuration\footnote{This ``odd'' symmetry is very approximate considering how far the current sheet deviates from the mid-plane at times. Nonetheless, we maintain this nomenclature to be consistent with the literature.}, with the toroidal field changing sign across the current sheet. In our other simulations, we generally find multiple co-existing current sheets (R5--b3p, R5--b4p--weak-cr), and current sheets which are off-mid-plane and not necessarily at a constant height over time (R5--b3p, R5--b4p--weak-cr, R5--b4n, all simulations with $R_{\rm AU} \geq 10$).

Our results are consistent with previous studies in that complex structures in $\langle B_y\rangle_{xy}$ are present regardless of field alignment or the inclusion of the Hall effect. For instance, \citet{Bai2013_wind1,Bai2013_wind2,Bai2013_mrioutflow} and \citet{Lesur2014} found off-mid-plane current sheets. \citet{Bai2015} often found off-mid-plane or multiple co-existing current sheets, and patterns where the sign of $\langle B_y \rangle_{xy}$ flips periodically for various radii and $\beta_0$.
\citet{Lesur2014} found that all of their simulations that include the Hall effect eventually eject the current sheet and exhibit even symmetry and speculate that all of their simulations are eventually symmetric on long enough timescales ($\gtrsim 10^3\;\Omega^{-1}$). However, our work and \citet{Bai2015} find that the system starting from an even configuration can occasionally spontaneously generate a current sheet and reverse the sign of $B_y$  (see \S\ref{sec:results:bfield_param}); thus, it is possible that current sheets will be present at later times in the disk evolution by spontaneous generation. Global disks do not possess the radial symmetry of the shearing box (i.e. the radial direction of the central star is known) and require an odd number of current sheets in or slightly vertically offset from the disk (see the simulations of e.g., \citealt{Bai2017,Bethune2017,Hu2023}) such that the toroidal magnetic field changes sign and the vertical gas outflow and magnetic field lines that leave the disk are bent radially away from the star both above and below the mid-plane.

The presence of a current sheet could have implications for the level of turbulence in our simulations. Although turbulence may be generated by magnetic reconnection via e.g. the tearing instability \citep{Galeev1976}, there are often only 2--4 grid cells across the width of a given current sheet for the simulations in our work. Because numerical simulations possess inherent numerical diffusion in addition to physical diffusion (e.g., Ohmic and ambipolar diffusion), it is not clear how strong a role physical diffusion processes have in affecting the current sheets in our simulations. With current sheet widths being close to the grid scale it seems likely that they are predominately affected by numerical diffusion and that any turbulence produced is the result of numerical reconnection and not small-scale instabilities.  However, we will still quantify the extent of turbulence produced by magnetic reconnection  in \S\ref{sec:results:bfield_param}, while keeping in mind its potential numerical origin.

\begin{figure}
    \centering
    \includegraphics[width=\columnwidth]{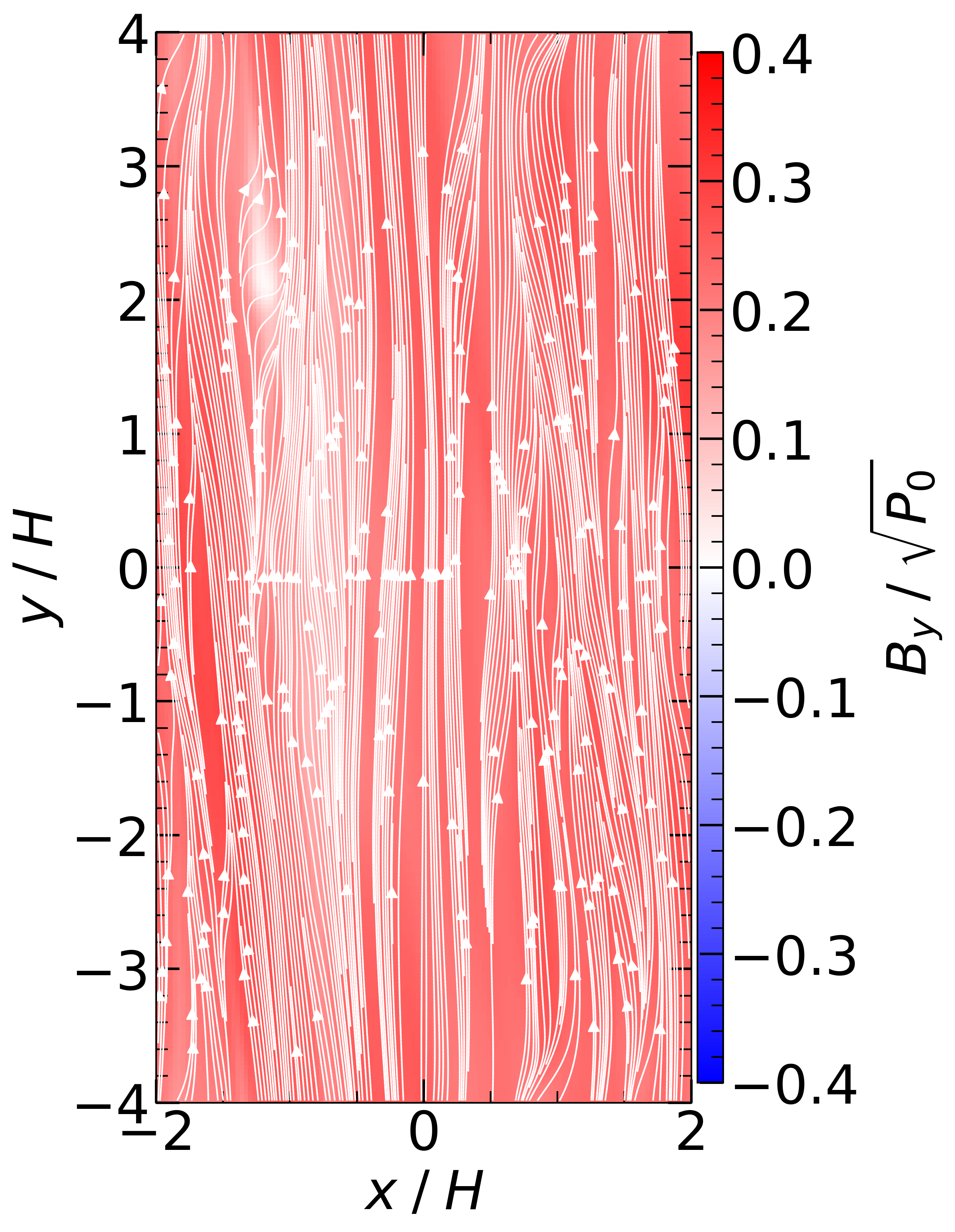}
    \caption{Slice in the $xy$-plane of the toroidal field $B_y(z=0)$ in the last snapshot of the 5 AU, $\beta_0=+10^4$ run. Streamlines are overlaid to show magnetic field lines. The field lines of $B_y$ are laminar, in that streamlines move in organized, parallel paths with minimal deviation.}
    \label{fig:bfield_xy}
\end{figure}

\begin{figure}
    \centering
    \includegraphics[width=\columnwidth]{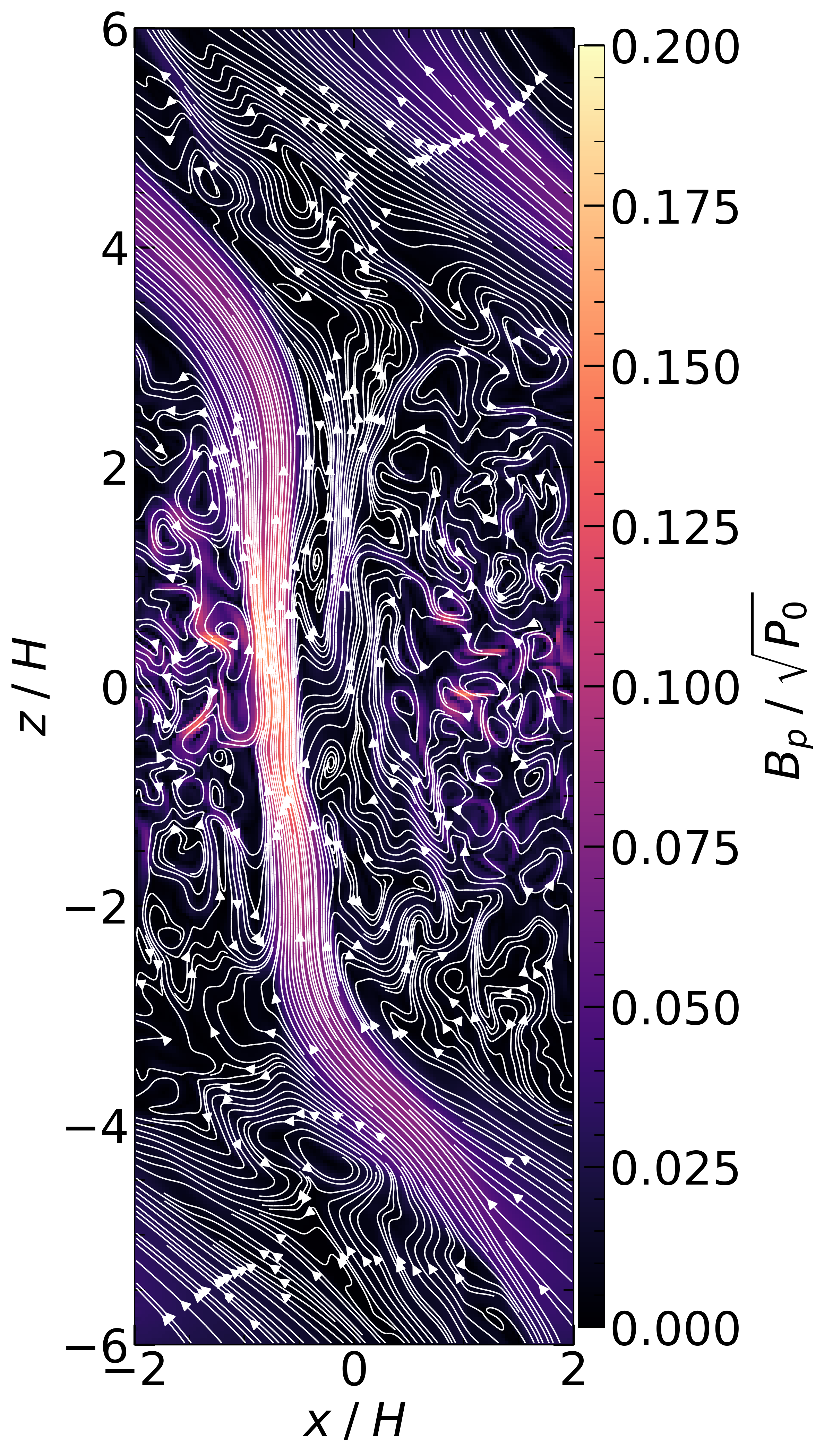}
    \caption{Slice in the $xz$-plane of the poloidal magnetic field $B_p(y=0)$ in the last snapshot of the 5 AU, $\beta_0=+10^4$ run. Streamlines are overlaid to show magnetic field lines. The poloidal field is much weaker than the toroidal field, except in a highly concentrated region of magnetic flux at $x\sim-0.5H$, where the strengths are comparable. Elsewhere in the shearing box, the poloidal magnetic field is weak but twisted and non-laminar.}
    \label{fig:bfield_xz}
\end{figure}

We examine the toroidal and poloidal magnetic field, in the last snapshot of the fiducial simulation (R5--beta4p; 5 AU, $\beta=+10^4$). Every other simulation is qualitatively similar. The magnetic field is dominated by the toroidal component (Figure \ref{fig:bfield_xy}). In a laminar velocity flow, fluid layers move smoothly past each other with little to no mixing; here the magnetic field lines are organized and nearly-parallel. Therefore, we characterize the magnetic field as laminar despite the field lines not representing a flow themselves. The poloidal field (Figure \ref{fig:bfield_xz}) has a single concentration of magnetic flux located at $x\sim-0.5 H$ that extends vertically for $\sim 3-4 H$ around the disk mid-plane. Magnetic organization of the vertical magnetic field is also found by \citet{Riols2019} with only Ohmic and ambipolar diffusion implemented. The poloidal field strength in this feature is comparable to that of the toroidal field, and appears alongside a weaker but more turbulent field with substantially tangled field lines; the convoluted poloidal magnetic field background may be an indicator of a turbulent flow despite a strong, laminar toroidal field.

\subsection{Velocity Structure}
\label{sec:results:velocity_structure}

Having made contact with the literature via an analysis of the magnetic field structure and evolution, we now turn to the question of the gas motions themselves. As mentioned previously, the velocity $\vec{v}$ is the gas velocity with the background shear component of the velocity $\vec{v}_0 = -q\Omega x \vec{\hat{e}}_y$ already removed.

We follow \citet{Simon2015_alma} in extracting the velocity fluctuations from the total velocity field at each snapshot. To remove bulk horizontal flows, we subtract the horizontally averaged velocities at each height:

\begin{align}
    \vec{v}' = \vec{v} - \langle\vec{v}\rangle_{xy}.
\end{align}

\noindent
We also remove any zonal flows -- large scale, axisymmetric radial variations in the azimuthal velocity -- by subtracting the azimuthal average of $v_y'$:

\begin{align}
    v_y'' = v_y' - \langle v_y'\rangle_y
\end{align}

\noindent
The turbulent velocity is then

\begin{align} \label{eq:deltav}
    \delta\vec{v} = \left(\delta v_x, \delta v_y, \delta v_z\right) = \left(v_x', v_y'', v_z'\right)
\end{align}

\begin{figure}
    \centering
    \includegraphics[width=\columnwidth]{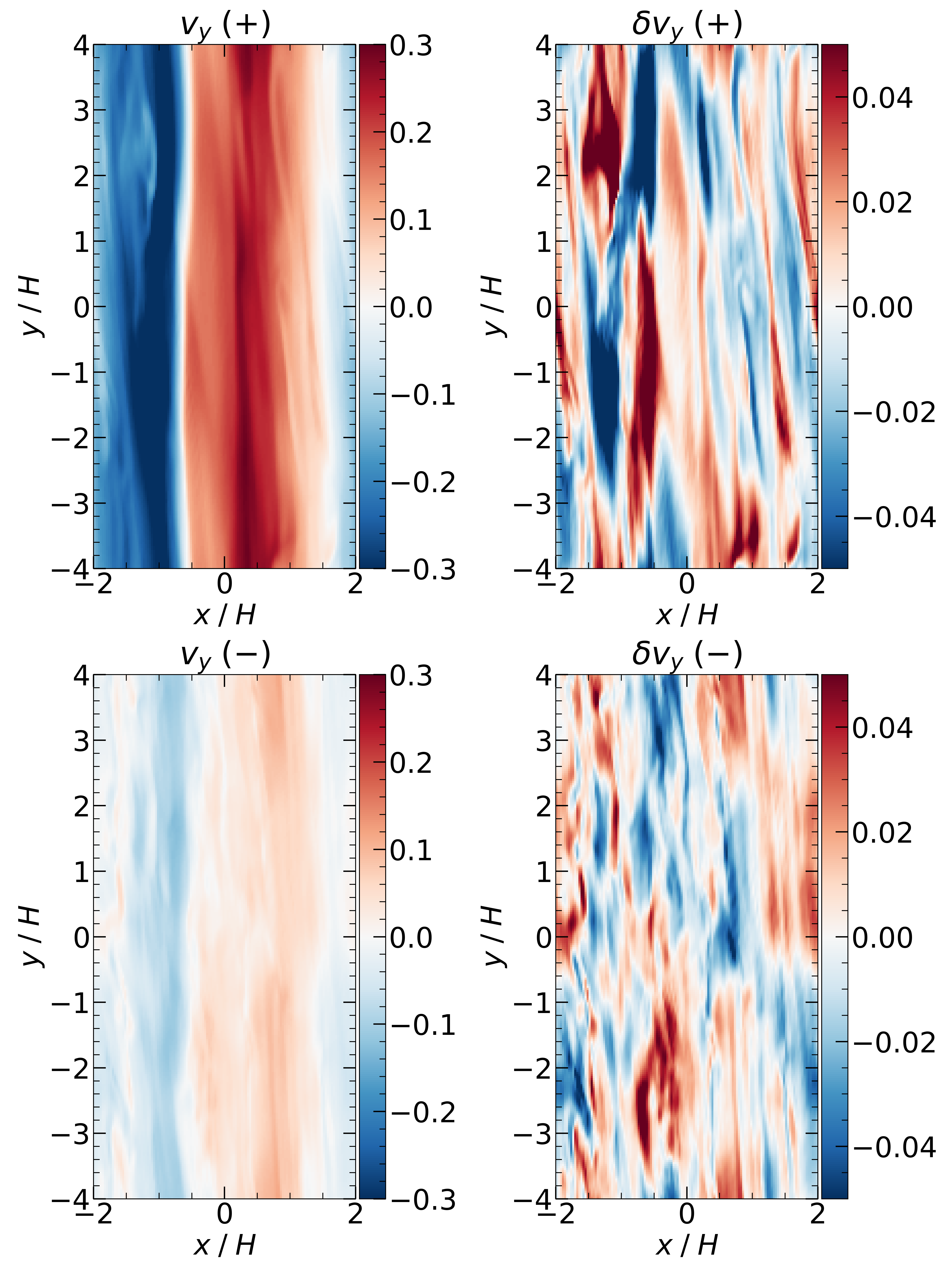}
    \caption{Azimuthal $(y)$ component of the mid-plane velocities for the simulations at 5 AU, with $\beta_0= +10^4$ (top) and $\beta_0 = -10^4$ (bottom). The total (though still shear-subtracted) velocity field (left panels) is dominated by large scale flows, the strength of which is sensitive to the polarity of the magnetic field. However, the turbulent velocity (right panels) is largely insensitive to the polarity. Note the different color scales. Some regions are purposely saturated in their color in order to emphasize the small scale turbulent modes.}
    \label{fig:midplane_velocities}
\end{figure}

Although this method produces velocity fluctuations that appear disordered, it does not guarantee that these fluctuations are turbulent; nonetheless, we refer to all three components of the velocity fluctuations as turbulent hereafter for simplicity.
We examine the velocities in the last snapshot for the simulations at 5 AU, $\beta_0=\pm 10^4$. The $y$-component of the mid-plane velocities are shown in Figure \ref{fig:midplane_velocities}. The total velocity field $v_y(\vec{x})$ (though still without the Keplerian shear flow) is dominated by a large scale, zonal flow like structure at a location which approximately matches that of the magnetic concentration in the poloidal field (Figure \ref{fig:bfield_xz}). The strength of this flow is sensitive to the polarity of the magnetic field, being stronger by a factor $\sim 4$ in the aligned case ($+$). However, the strength of the small scale fluctuations $\delta v(\vec{x})$ is not sensitive to the magnetic field polarity.

\begin{figure}
    \centering
    \includegraphics[width=\columnwidth]{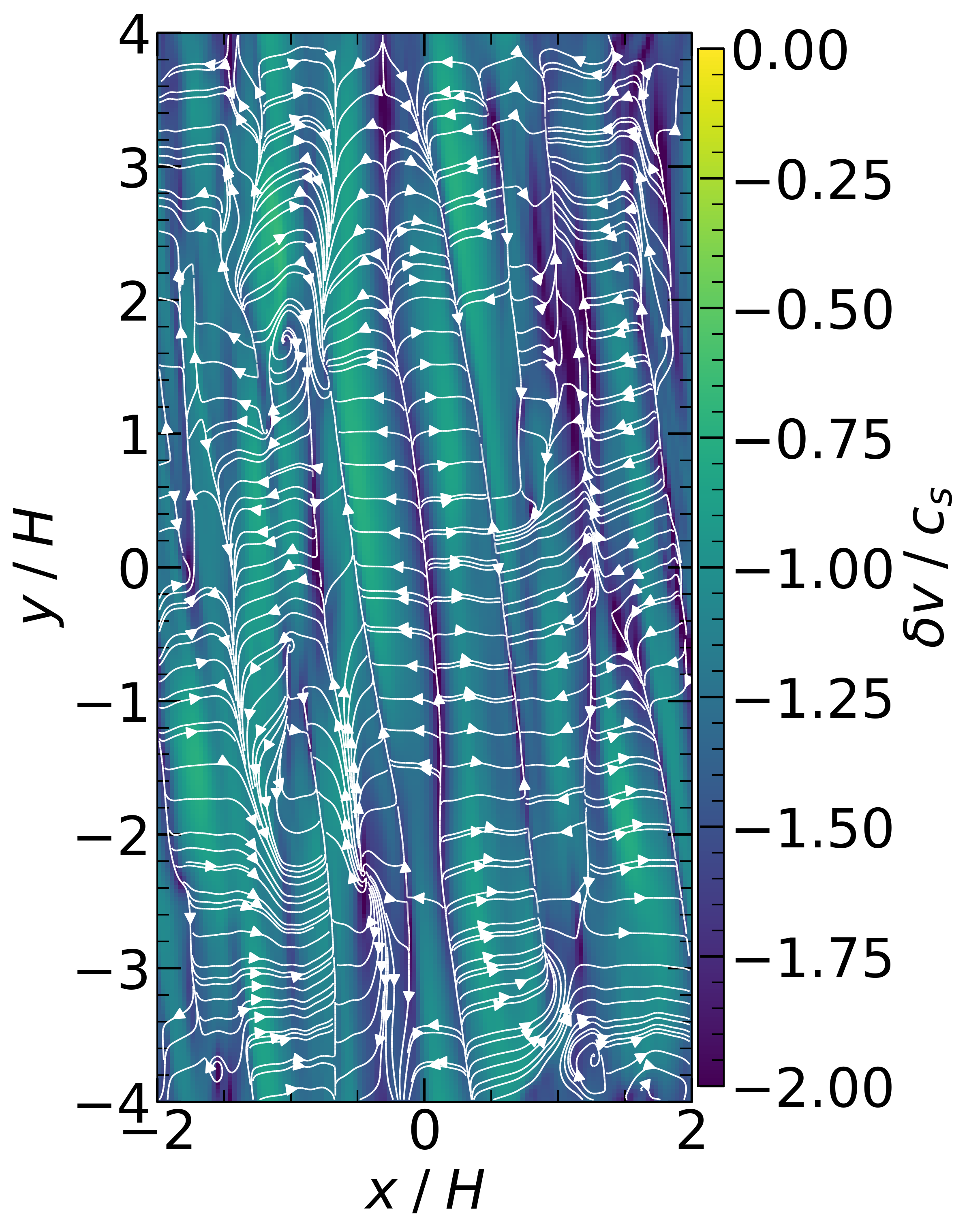}
    \caption{The $xy$-slice of the turbulent velocity $\delta v(z=0)$ in the last snapshot of the 5 AU, $\beta_0=+10^4$ run. Streamlines of $\delta \vec{v}$ (Equation \ref{eq:deltav}) show deviations from the mean flow. The flows in the $xy$-plane do not appear to be strongly turbulent (i.e., the velocity structure appears somewhat ordered),  possibly because fluctuations in this plane are stretched by the background shear.}
    \label{fig:turbvel_xy}
\end{figure}

\begin{figure}
    \centering
    \includegraphics[width=\columnwidth]{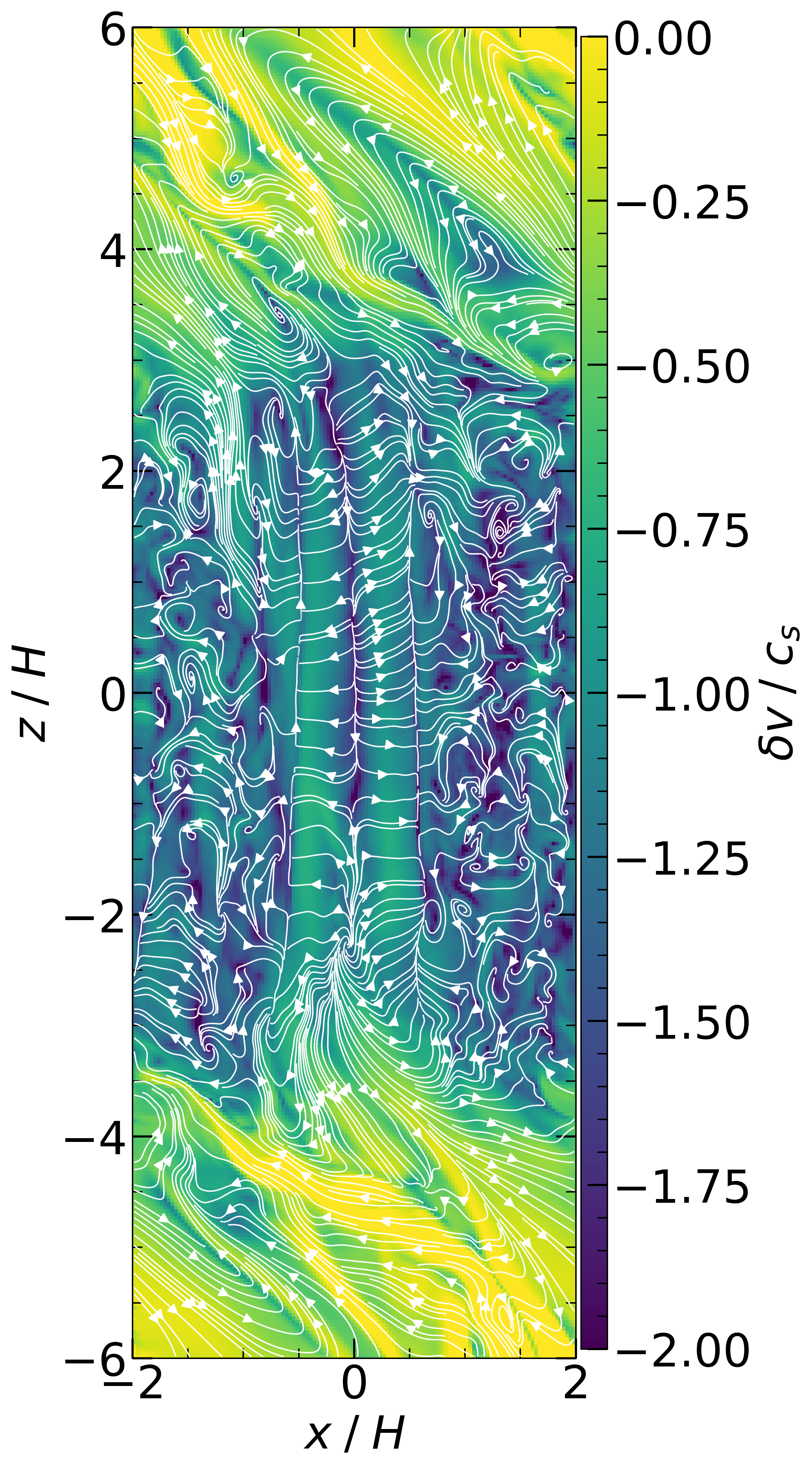}
    \caption{The $xz$-slice of the turbulent velocity $\delta v(y=0)$ in the last snapshot of the 5 AU, $\beta_0=+10^4$ run. Streamlines of $\delta \vec{v}$ (Equation \ref{eq:deltav}) are also shown. The gas appears to be turbulent in the $xz$-plane, with a large number of eddies and a disorganized flow.}
    \label{fig:turbvel_xz}
\end{figure}

\begin{figure}
    \centering
    \includegraphics[width=\columnwidth]{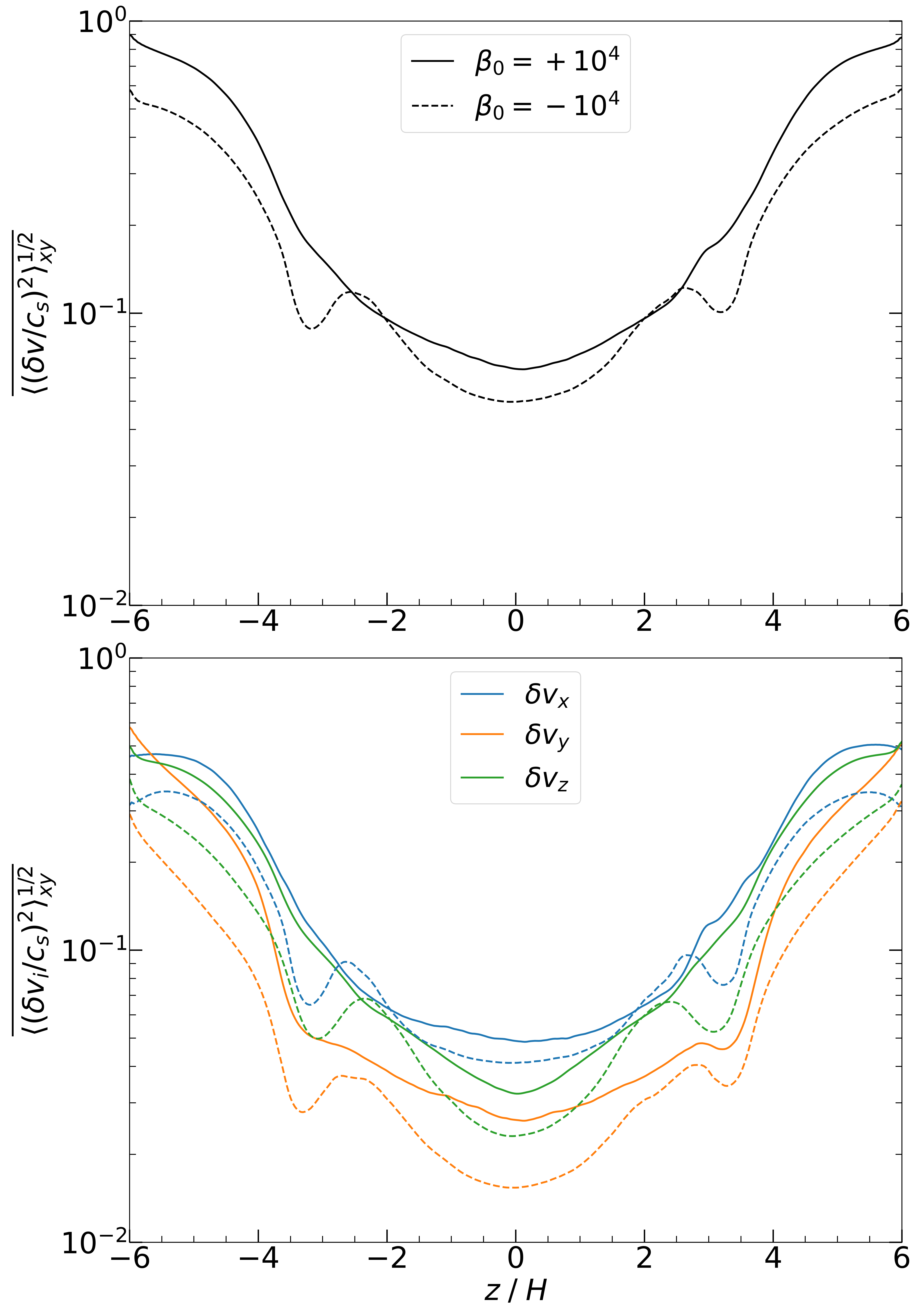}
    \caption{Time-averaged and and RMS turbulent velocity profile normalized by the sound speed vs. vertical height. The turbulent velocity is large ($\sim c_s$) in the highly ionized upper layers of the disk. In the mid-plane, the turbulent velocities are weaker but still substantial, with $\delta v_{|z|<H} \sim 5$--$7\times 10^{-2}$. The turbulent velocity components are consistently within a factor $\sim 3$ of each other.}
    \label{fig:velocity_profile}
\end{figure}

The flows that exist in the mid-plane (Figure \ref{fig:turbvel_xy}) appear to be somewhat organized, with only a few apparent eddies. However, the velocity fluctuations in the vertical plane (Figure \ref{fig:turbvel_xz}) do appear to be turbulent, with many eddies and stochastic flow. The level of the turbulence has a strong dependence on height in part due to the highly variable disk ionization (as in e.g. \citealt{Bai2015,Simon2018}) and the strength of $\beta$ (e.g. see Figure 2 of \citealt{Yang2018}). Figure \ref{fig:velocity_profile} shows the magnitude and components of the time- and horizontally-RMS-averaged turbulent velocities. Near the disk surface, the turbulent velocity magnitudes approach the sound speed $c_s$ in agreement with a number of other works that explored a large range of parameters and numerical setups (e.g., \citealt{Fromang2009,Simon2011_stratified,Simon2011_linewidths,Flock2012,Simon2015_alma}); evidently, this increase in turbulent velocity up to approximately the sound speed is a robust feature of magnetized disks for a number of different parameters. We further quantify the mid-plane velocities by averaging $\overline{\langle\delta v^2\rangle_{xy}^{1/2}}$ over the innermost two scale heights, i.e. for $|z| < H$ (Table \ref{table:simulations}). At 5 AU, the mid-plane turbulence is $\approx 6.8\times 10^{-2}c_s$ when $\beta_0 > 0$ and $\approx 5.2\times 10^{-2}c_s$ when $\beta_0 < 0$. In general, the strength of the mid-plane turbulence in our simulations differs by only a factor $< 2$ between the corresponding $\beta_0 > 0$ and $\beta_0 < 0$ configurations. The magnitude of the components of $\langle\delta v\rangle_{xy}$ are each within a factor $\sim 3$ of the others. The vertical component of turbulent velocities was briefly examined by \citet{Bai2015}; those turbulent velocities near the midplane are $\sim 2$--$3$ times lower than that of the corresponding runs in our simulations, but \citet{Bai2015} considered a cosmic ray ionization rate of $10^{-17}~{\rm s^{-1}}$, an order of magnitude lower than our fiducial value.

\subsection{Turbulent Velocity Power Spectrum}
\label{sec:results:power_spectrum}

Gas turbulence is often described as being composed of eddies having a range of spatial scales $l$ and spatial frequencies $k =  2\pi/l$. The energy spectrum is a measure of the energy contained at a frequency $k$. The Fourier transform for some flow quantity $Q$ is defined as

\begin{equation}
    \widetilde{Q}(\vec{k}) = \int Q(\vec{x})e^{-i\vec{k}\cdot\vec{x}} dx^3
\end{equation}

We examine the turbulent kinetic energy in Fourier space, defined as

\begin{equation}
    E_{\rm 3D}(\vec{k}) = \frac{1}{2} \sum_{i=x,y,z}\left| \widetilde{\sqrt{\rho_0}\delta v_i}(\vec{k}) \right|^2.
\end{equation}

\begin{figure}
    \centering
    \includegraphics[width=\columnwidth]{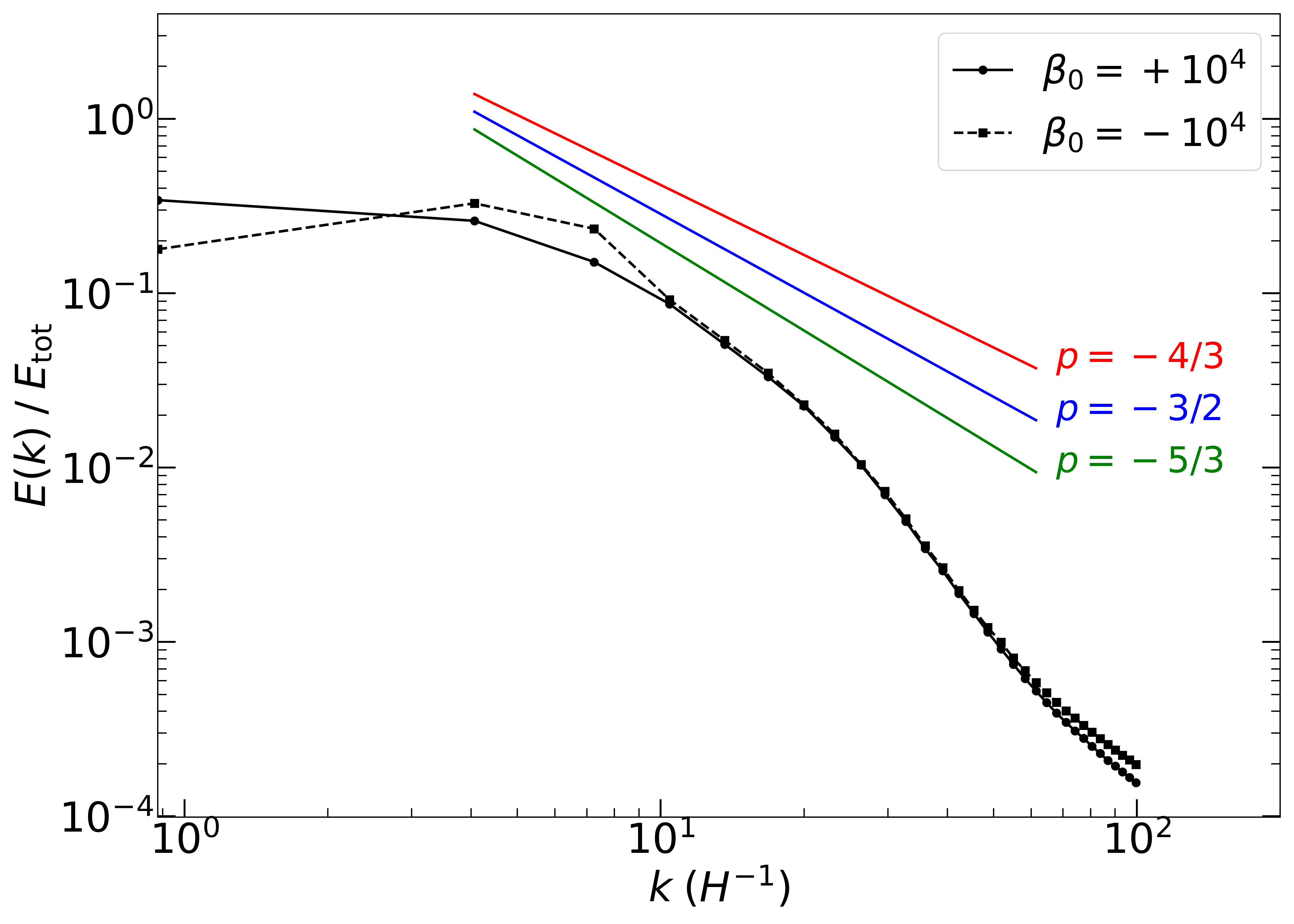}
    \caption{Turbulent kinetic energy spectra at 5 AU, for $\beta_0 = \pm 10^4$. Neither spectra exhibit a clear inertial range with a power law scaling $k^{p}$. The slopes are somewhat steeper than $p=-4/3$ or $p=-3/2$, which have been reported from previous studies of MHD turbulence (see main text).}
    \label{fig:energy_spectra}
\end{figure}

\noindent
We then integrate over spherical shells of radius $k~=~|\vec{k}|~=~\sqrt{k_x^2 + k_y^2 + k_z^2}$ and width $dk$,

\begin{align}
    E(k)dk &= E_{\rm 3D}(\vec{k}) dk^3 \nonumber \\
    E(k) &= \int_{k-dk/2}^{k+dk/2} E_{\rm 3D}(\vec{k'}) 4\pi k'^2 dk'
\end{align}

\noindent
and normalize by the total energy $E_{\rm tot} = \int E(k)\;dk$. Here, $E_{\rm 3D}(\vec{k})$ denotes the three-dimensional spectrum obtained via the Fourier transform, and $E(k)$ is the one-dimensional energy spectrum obtained by integrating $E_{\rm 3D}(\vec{k})$. The resulting energy spectra are shown in Figure \ref{fig:energy_spectra}.

The spectra broadly resembles that of an energy cascade from small wavenumbers, where energy is introduced into the system at large scales, to large wavenumbers, where energy is dissipated by viscous and magnetic diffusion. However, the energy spectra noticeably lacks a distinct inertial range at intermediate wavenumber (similar to e.g. \citealt{Fromang2007,Simon2009_turb}), which should follow a simple power law $E(k) \propto k^{p}$.

The precise form of the energy spectrum has important implications for planet formation, as turbulence is the dominant force driving collisions between grains \citep{Ormel2007}. The collision velocities between small grains are extremely sensitive to $p$, and can vary by up to an order of magnitude for an inertial range scaling $-5/3 < p < -3/2$ \citep{Gong2021}. Many previous numerical MHD simulations have reported either $p \sim -4/3$ for driven MHD turbulence \citep{Lemaster2009,Salvesen2014,Makwana2015,Grete2017,Gong2020} or $p \sim -1.5$ for turbulence driven by the MRI \citep{Fromang2010,Lesur2011,Walker2016}. Only a small number of the works referenced above that report $p=-4/3$ consider non-ideal MHD (these being \citealt{Lesur2011,Salvesen2014,Walker2016}) and these works only consider Ohmic diffusion. Additionally, many of the above works also contain spectra with a small or non-existent\footnote{Our preference for a definitive inertial range would be a clear power law scaling that lasts well over one decade in $k$, which is not typically seen in these previous works (but see e.g. \citealt{Salvesen2014}.} inertial range akin to the spectra in Figure \ref{fig:energy_spectra}.

\begin{figure}
    \centering
    \includegraphics[width=\columnwidth]{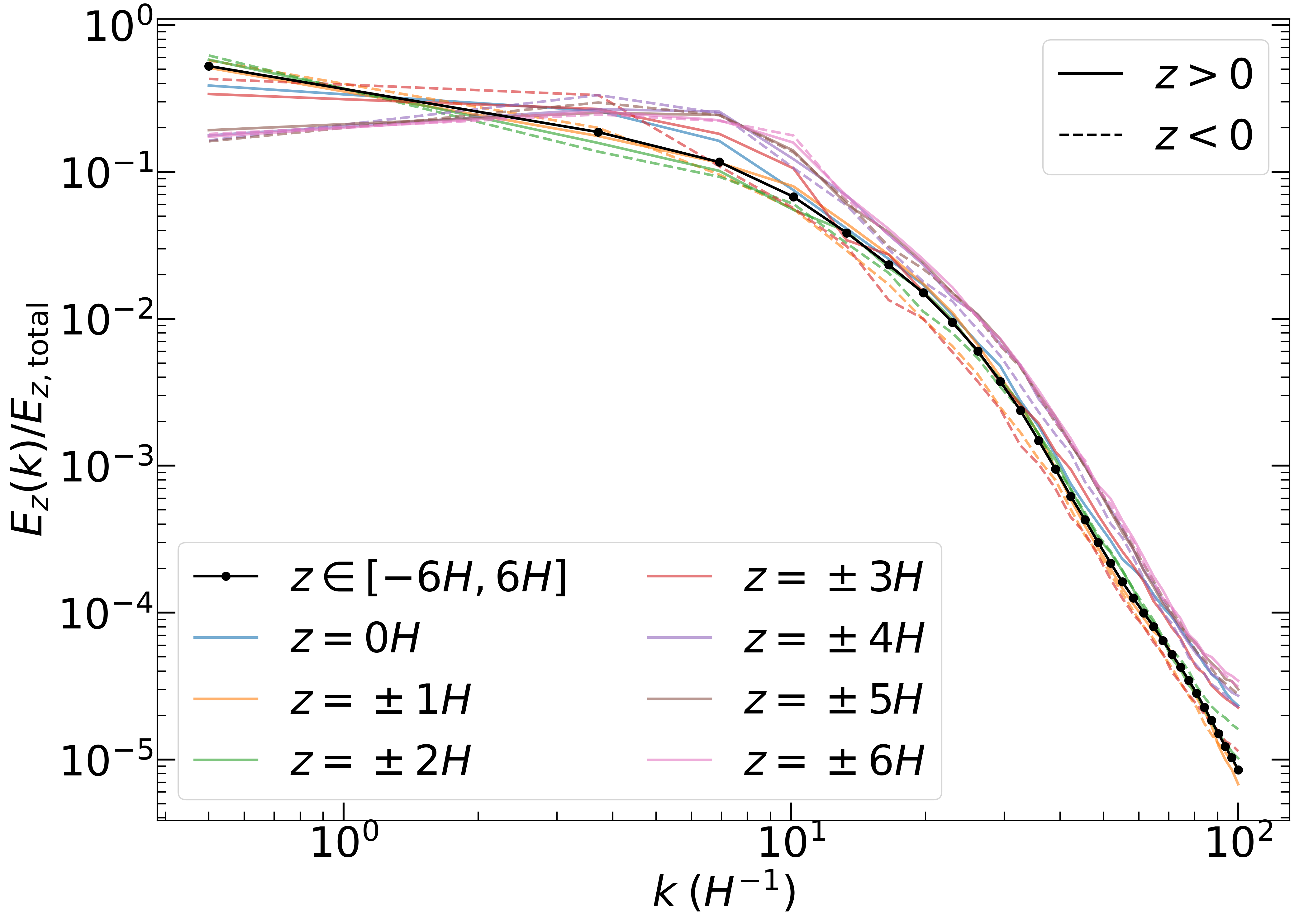}
    \caption{Energy spectra for the $\beta_0 > 0$ case, computed at different heights above (solid) and below (dashed) the disk mid-plane, averaged for the last $40~\Omega^{-1}$ of the simulation. The spectra have similar scaling (and no inertial range) despite the large variation in turbulence across height.}
    \label{fig:spectra_across_height}
\end{figure}

Because the magnitude of turbulence has a strong dependence on the vertical distance from the mid-plane --- low-Mach number turbulence located in the mid-plane region ($|z|~\lesssim~3H$), and the relatively large-Mach number turbulence located near the disk surface ($|z|~\gtrsim~3H$) --- we quantify the effect of this height dependence on the energy spectra. We computed spectra for the $+\beta_0$ case at different heights above (and below) the mid-plane, averaged over the last $40~\Omega^{-1}$ of the simulation (Figure \ref{fig:spectra_across_height}). Here $E(k)$ is obtained by integrating the 2D spectra (at some height $z$) over annuli of radius $k = \sqrt{k_x^2 + k_y^2}$ and width $dk$. Despite the large variation in the turbulent strength across height, each spectra has mostly similar scaling and no clear inertial range, though at large vertical distances from the mid-plane ($z \gtrsim 4H$) the slope is slightly shallower at small wavenumber. Since there remains no clear inertial range at these separate heights, we conclude that the lack of an inertial range in our energy spectra is not the result of computing the spectra for a mixture of low-Mach and high-Mach turbulence and is thus more likely to be a result of insufficient resolution. However, it remains unclear what resolution would be necessary to see a clear inertial range.

\subsection{Turbulent Velocity Autocorrelation Function}
\label{sec:results:acf}

\begin{figure*}
    \centering
    \includegraphics[width=\linewidth]{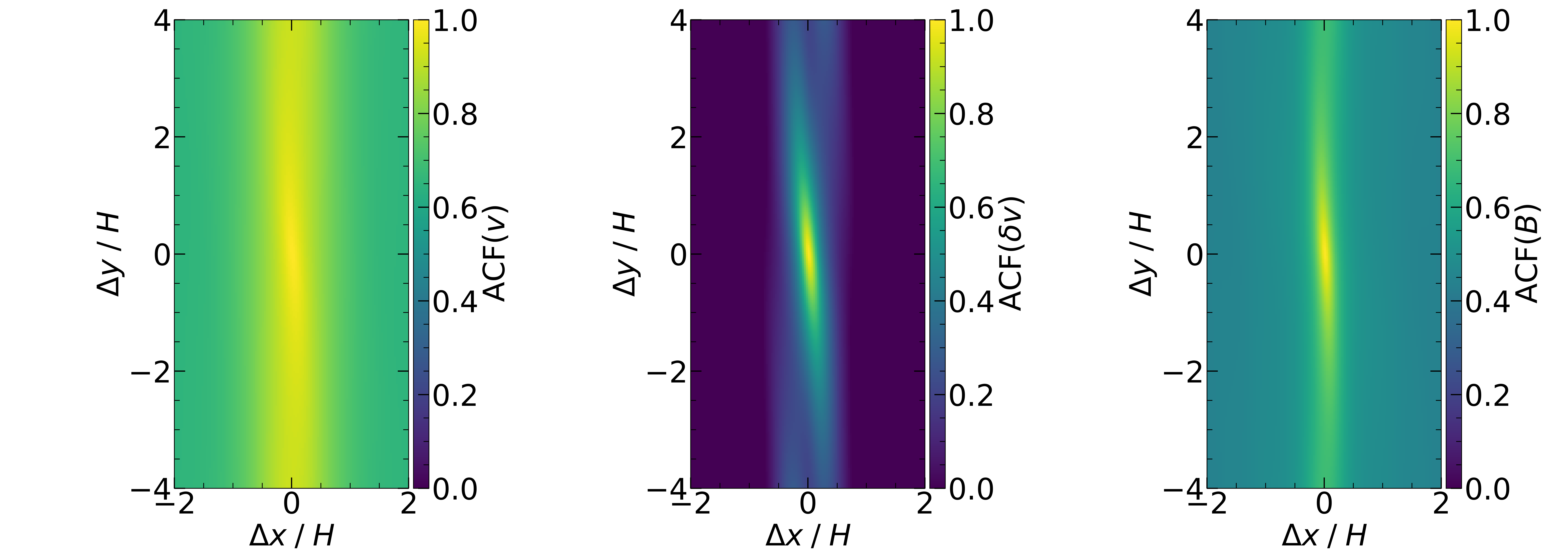}
    \caption{Time averaged autocorrelation functions (ACF) at zero offset from the vertical ($\Delta z = 0$) of the total velocity field (left), turbulent velocity field (center), and magnetic field (right).}
    \label{fig:acf}
\end{figure*}

We also examine the spatial autocorrelation function (ACF; \citealt{Guan2009,Simon2012,Simon2013_weakacc,Bai2013_mrioutflow}) to examine structure in the $xy$-plane,

\begin{align}
    {\rm ACF}(Q(\Delta\vec{x})) \equiv \overline{\left[ \frac{\int Q(t,\vec{x})\;Q(t,\vec{x}+\Delta\vec{x})\;d^3\vec{x}}{\int Q(t,\vec{x})^2\;d^3\vec{x}} \right]}
\end{align}

\noindent
where $Q$ is the quantity of interest. For vector quantities, we follow \citet{Guan2009} and \citet{Simon2012} and define ${\rm ACF}(\vec{Q}) = \sum_{i=x,y,z} {\rm ACF}(Q_i)$. We consider the ACF at a vertical offset $\Delta z = 0$. The autocorrelation functions (Figure \ref{fig:acf}) show two distinct components, consistent with previous studies: a strong, tilted centroid region, and a volume-filling background component. The centroid region denotes a strong, localized correlation in the field, while the background component corresponds to the velocity flow (and magnetic field) structures on the largest scale of the numerical domain \citep{Simon2012}. The ACFs of the total velocity and magnetic field contain strong background components; this indicates strong, large scale  structures in these fields. As expected, the ACF of the turbulent velocity does not contain this background component ($|{\rm ACF}(\delta v)| < 0.2$ for most of the domain). The centroid tilt angle from the vertical axis is empirically related to the Maxwell and Reynolds components of the stress parameter $\alpha$ \citep{Guan2009}. The ACF$(\delta v)$ in this work displays a tilt angle $\sim 6^\circ$, in good agreement with the turbulence found in \citet{Guan2009}. Studies of MRI-driven turbulence in ideal MHD \citep{Guan2009,Beckwith2011,Simon2012,Bai2013_mrioutflow} find centroid tilt angles that agree with the ratio of Maxwell stress to $B^2$ ($\theta_B \approx \theta_{\rm tilt} \approx 15^\circ$, where $\langle B_xB_y\rangle/\langle B^2\rangle = \sin{2\theta_B}$). In our work, ACF($B$) exhibits a tilt angle $\sim 2.5^\circ$ in our simulations, which is smaller than predicted by the characteristic angle $\theta_B$, which we compute to be $\sim 5^\circ$. Such a small tilt angle (compared with MRI driven turbulence) seems to be reflective of the degree to which the laminar toroidal field dominates over the other magnetic field components. For example, in the ideal MRI simulations of \citet{Simon2009_turb} $\langle B_y^2\rangle/\langle B_x^2\rangle \sim 10$, while in our simulations $\langle B_y^2\rangle/\langle B_x^2\rangle$ is more typically $\sim 100$. However, the cause of disagreement between $\theta_B$ and $\theta_{\rm tilt}$ is not clear. The small tilt angles are also consistent with the elongated azimuthal structure of the turbulence (see Figure \ref{fig:midplane_velocities}) compared to the radial structure. 

While the centroid is well contained by the simulation domain in the turbulent velocity autocorrelation, we note the presence of a tail along the centroid major axis that wraps around through the periodic boundary in azimuth into the other side of the domain. This implies that there may be flows which are correlated in azimuth on scales larger than the numerical domain. Referring again to Figure \ref{fig:midplane_velocities}, the turbulence consists of small length scales in the $x$-direction, but larger length scales (possibly larger than the azimuthal domain, particularly for the $\beta_0 > 0$ case) in the $y$-direction. While a larger azimuthal extent would better capture the tail of the centroid, the ACF$(\delta v)$ along the major axis tail of the centroid is $\lesssim 0.3$ at the boundary of the domain; thus, the main centroid of strong correlation is still well-contained within the simulation domain.

\subsection{Vertical Diffusion}
\label{sec:results:diffusion_param}

We follow \cite{Zhu2015} and \cite{Youdin2007} to compute an estimate for the particle scale height $H_d$. We compute the vertical gas diffusion coefficient via

\begin{align}
    D_{g,z} = \int_0^\infty \langle v_{g,z}(t)v_{g,z}(0)\rangle_{|z|<H}\;dt
\end{align}

\noindent
where $v_{g,z}$ is the vertical velocity of the gas. We find $D_{g,z} \approx 1.0\times 10^{-3}$, which is roughly consistent with previous numerical simulations with large resistivity \citep{Gole2016}.

\begin{figure}
    \centering
    \includegraphics[width=\columnwidth]{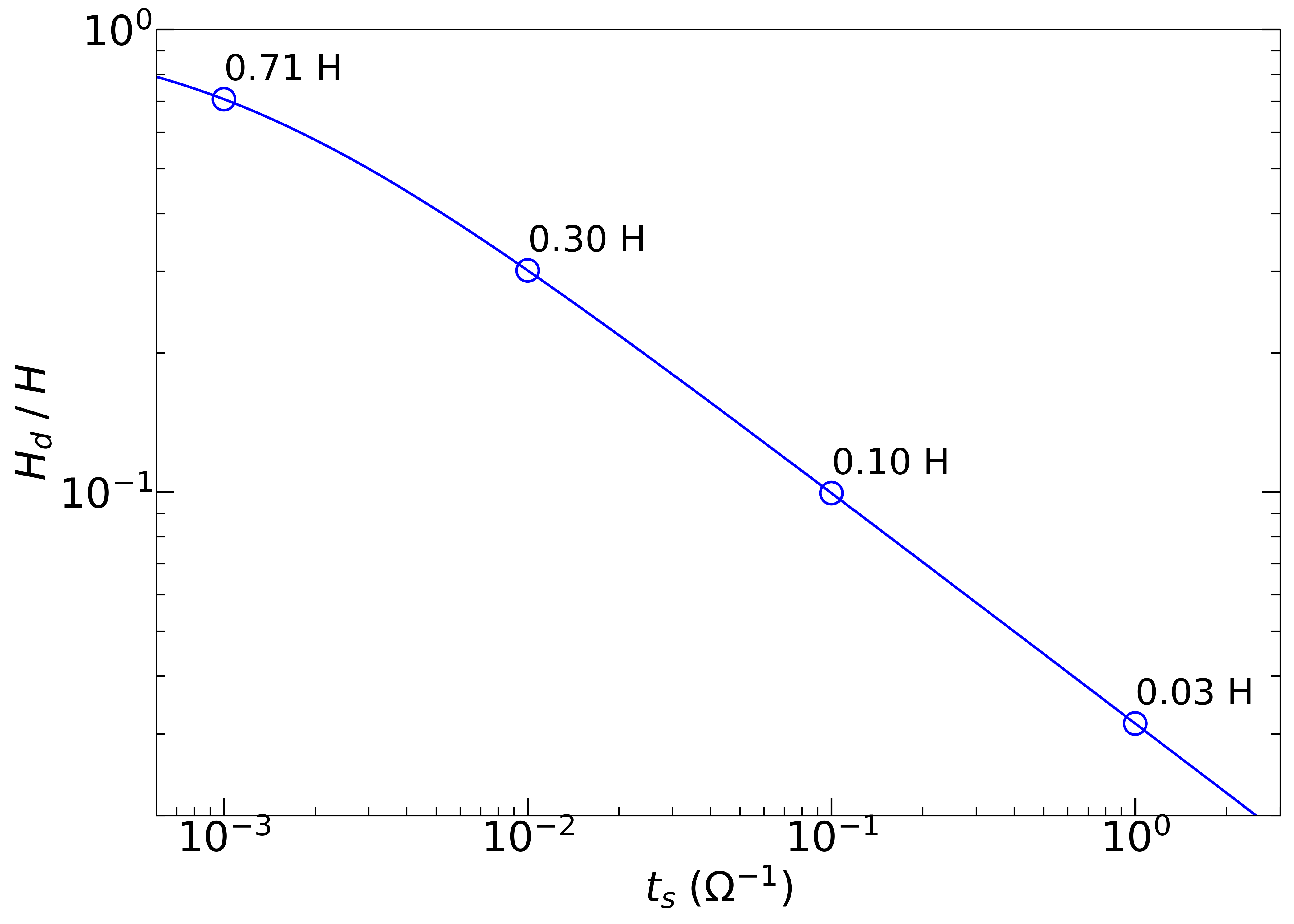}
    \caption{Particle scale height $H_d$ as a fraction of the gas scale height $H$ (from Equation \ref{eq:diffusion}) vs. stopping time in units of $\Omega^{-1}$ given the vertical gas diffusion coefficient $D_{g,z} = 1.0\times 10^{-3}$. Circles mark the particle scale height at each decade. Large particles are more settled, but smaller particles may be diffused to a significant fraction of the gas scale height.}
    \label{fig:diffusion}
\end{figure}

The particle scale height (neglecting particle mass, and thus momentum feedback), is \citep{Dubrulle1995,Youdin2007,Zhu2015}

\begin{align} \label{eq:diffusion}
    H_d = \frac{H}{\sqrt{c_s^2 t_s D_{g,z}^{-1} + 1}}
\end{align}

\noindent
where $t_s$ is the stopping time of the particle. The stopping time is the time over which a particle velocity is reduced by a factor $e$ by a constant drag force (e.g. gas headwind) and is proportional to the particle size in the Epstein drag regime \citep{Weidenschilling1977}. Because the energy spectra and overall level of mid-plane turbulence is similar for both aligned and anti-aligned magnetic fields, we only examine the vertical stirring for $\beta_0 > 0$; the results are shown in Figure~\ref{fig:diffusion}. The vertical velocities can potentially diffuse small particles with $t_s\Omega = 10^{-3}$ (important for ionization, see \S\ref{sec:discussion:uncertainties:dust_grains}) to $H_d = 0.71\;H$, a significant fraction of the gas scale height. Particles with $t_s\Omega = 0.1$ can be diffused to $H_d \sim 0.1\;H$ (Figure \ref{fig:diffusion}). This is approximately the same vertical diffusion as found for an Ohmic dead zone with dust-to-gas mas ratio $Z = 2$--4\% \citep{Yang2018} and disks with ambipolar diffusion (and multiple particle sizes, \citealt{Zhu2015}), but is much smaller than what is possible for the hydrodynamic vertical shear instability ($H_d \sim 0.5 H$ for $t_s\Omega = 0.1$; \citealt{Dullemond2022}). However, including the influence of particle momentum feedback on the gas (i.e., giving the particles non-negligible mass) can result in a much more settled disk, with $H_d/H \sim 0.02$--0.04 for $Z = 1$--2\% (\citealt{Xu2022}; see also \citealt{Lim2023_arxiv}); as the current simulations do not contain particles, the values of $H_d$ in Figure \ref{fig:diffusion} should be taken as upper limits.

\subsection{Effects of Magnetic Field Strength and Current Sheet}
\label{sec:results:bfield_param}

\begin{figure}
    \centering
    \includegraphics[width=\columnwidth]{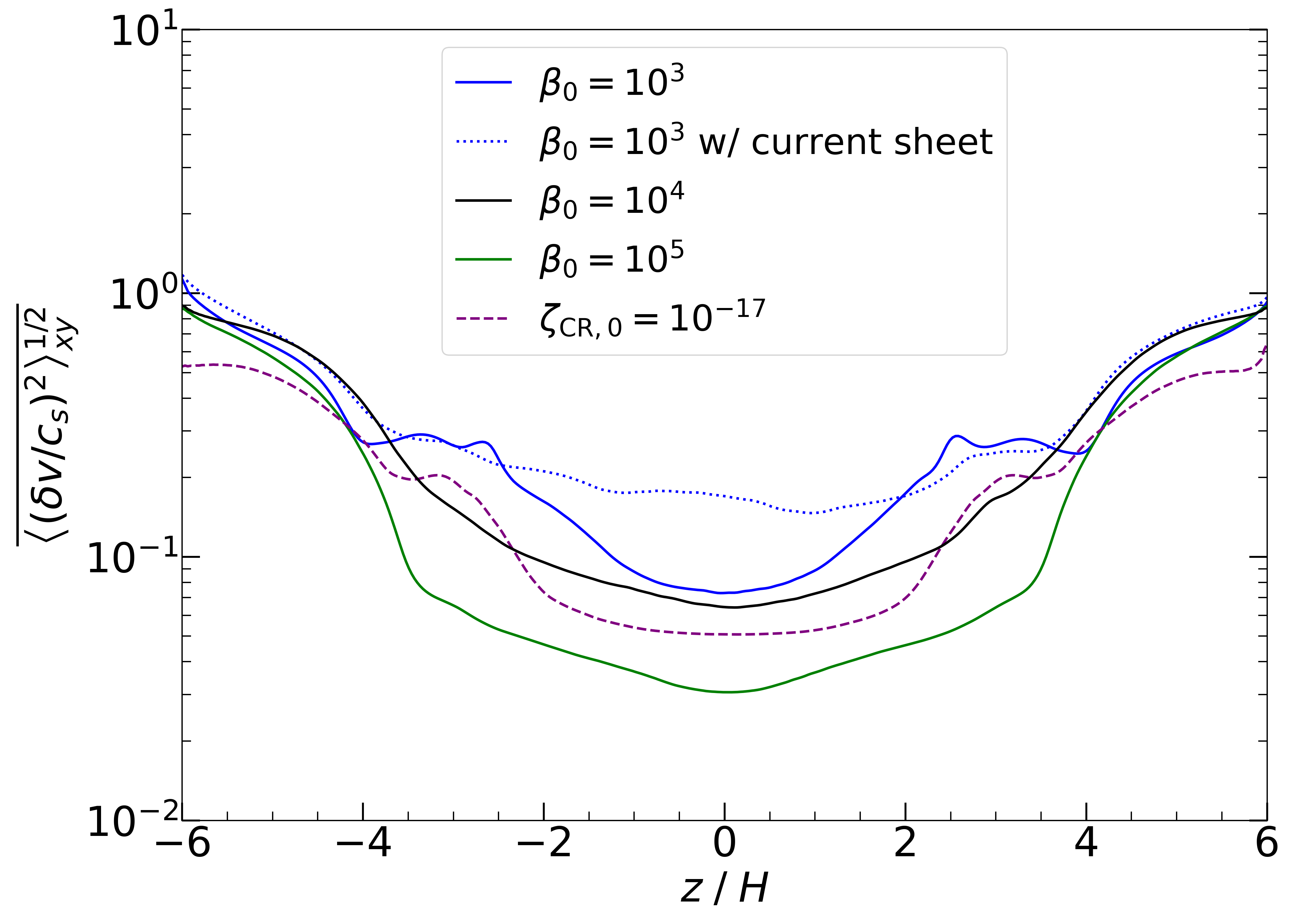}
    \caption{Same as Figure \ref{fig:velocity_profile}, but varying the strength of the magnetic field (i.e., varying $\beta_0$). A stronger (weaker) magnetic field results in stronger (weaker) gas turbulence. In the strong field simulation ($\beta_0 = +10^3$) the presence of a current sheet is correlated with enhanced mid-plane gas turbulence. Reducing the cosmic ray flux ($\zeta_{\rm CR} = 10^{-17}$) does not significantly reduce the level of mid-plane gas turbulence. }
    \label{fig:bfield_param_deltav}
\end{figure}

We investigate the effect of different initial magnetic field strengths on the turbulent velocities at 5 AU (Figure \ref{fig:bfield_param_deltav}). For a weak initial magnetic field ($\beta_0 = +10^5$), $\overline{\alpha}~\sim~3~\times~10^{-4}$. The toroidal field exhibits an initial transient current sheet, but afterward maintains an even symmetry for the duration of the simulation. The time-averaged turbulent velocity magnitude in the mid-plane region (which we define to be the innermost two scale heights, or $|z|~<~H$) is $\sim~3.3\times 10^{-2}~c_s$, smaller by a factor $\sim 2$ compared with the $\beta_0 = +10^4$ configuration. These time-averages correspond to times when $\langle B_y\rangle_{xy}$ does not exhibit a current sheet.

\begin{figure*}
    \centering
    \includegraphics[width=\textwidth]{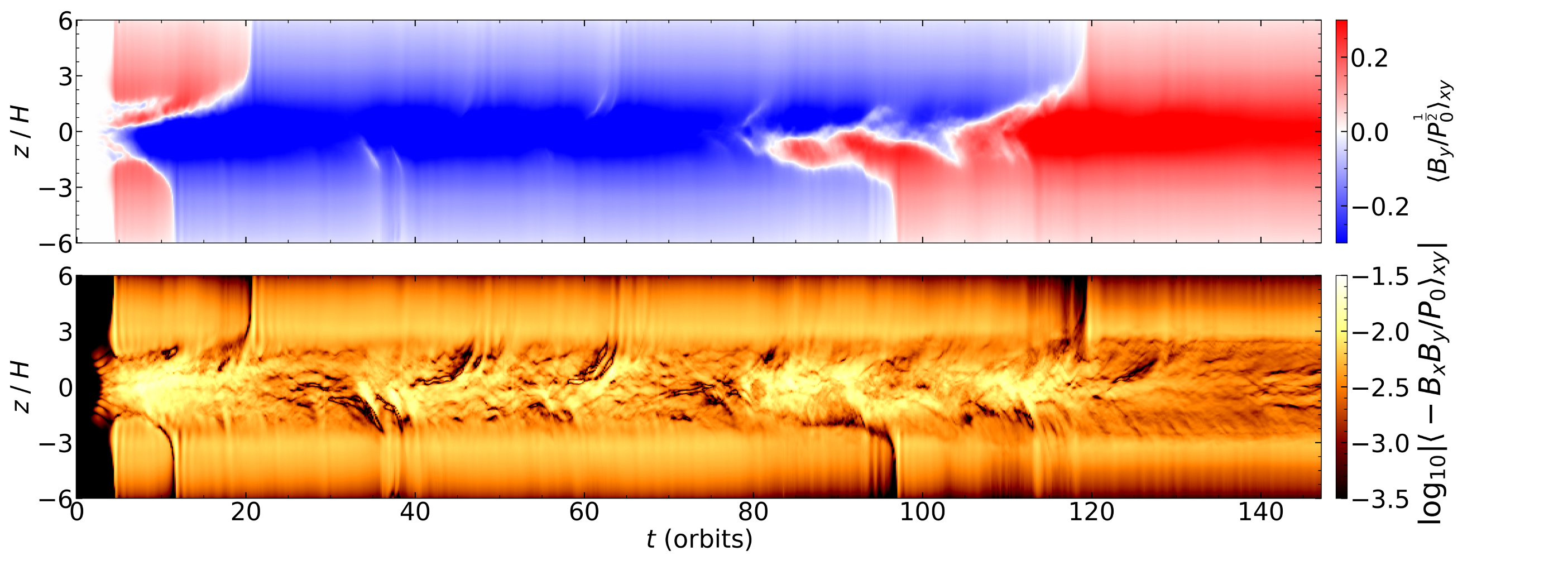}
    \caption{Spacetime diagrams of the toroidal field $\langle B_y\rangle_{xy}$ and Maxwell stress $\log_{10}\langle -B_xB_y\rangle_{xy}$ for the simulation at 5AU with $\beta_0=+10^3$. In the toroidal field, current sheets are spontaneously generated at around $t=80$ orbits, both of which are ejected from the simulation domain by 120 orbits. The times at which these current sheets exist are correlated with times of enhanced mid-plane turbulence.}
    \label{fig:st_5AU_beta3}
\end{figure*}

\begin{figure}
    \centering
    \includegraphics[width=\columnwidth]{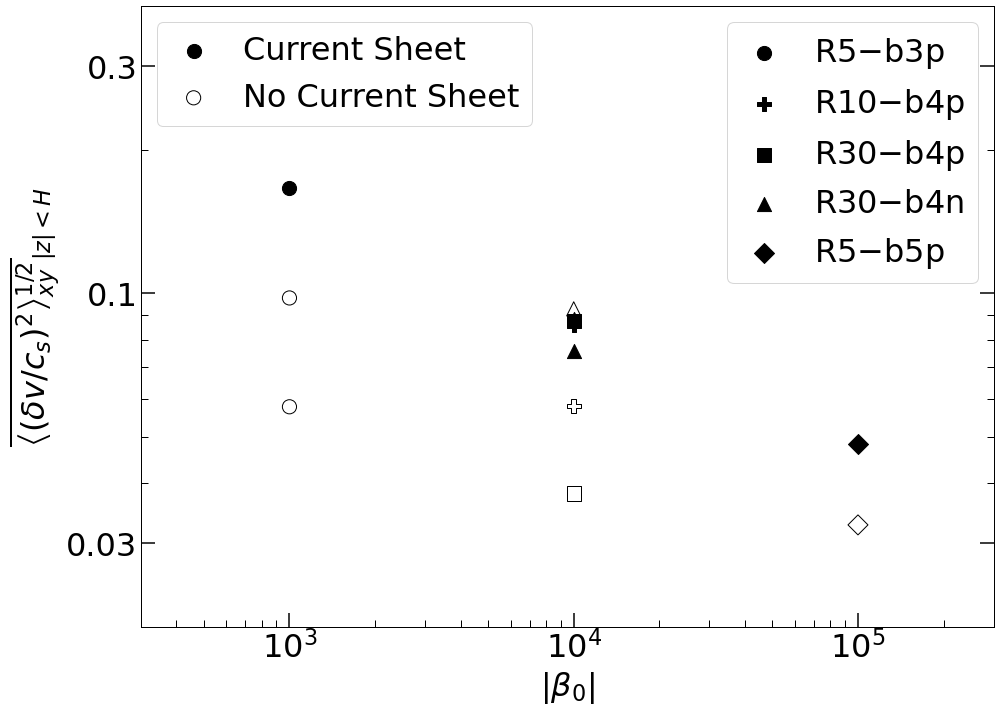}
    \caption{Mid-plane turbulent velocities averaged over times with (filled) and without (unfilled) current sheets exhibited by $\langle B_y\rangle_{xy}$. Different shapes correspond to different initial configurations in $\beta_0$ and radius. In general, the mid-plane turbulence is enhanced by a factor $\sim$1.5--3 when a current sheet is present.}
    \label{fig:cs_comparison}
\end{figure}

For a strong initial magnetic field ($\beta_0 = +10^3$), $\overline{\alpha}~\sim~5~\times~10^{-3}$, but the mid-plane turbulent velocity magnitudes are not significantly larger than the $\beta_0 = +10^4$ configuration. Looking more closely at the magnetic field geometry, we note the strong field case results in a toroidal field with even symmetry (no current sheet) after initial transients until $t \approx 80$ orbits, at which time two current sheets are spontaneously generated near the mid-plane (Figure \ref{fig:st_5AU_beta3}). At least one current sheet lasts for the next $\sim 40$ orbits, after which it is ejected from the shearing box and the sign of $\langle B_y\rangle_{xy}$ changes. Similar spontaneous current sheet generation was seen by \citet{Bai2015}, although it is not understood why this occurs. In the range of times over which a current sheet exists, the turbulent velocities at the mid-plane average to $\sim~1.7~\times~10^{-1}~c_s$, which is enhanced by a factor $\sim 2.5$ times greater than the $\beta_0 = 10^3$ configuration without a current sheet. This is a general behavior of our simulations; Figure \ref{fig:cs_comparison} shows the differences in mid-plane turbulent velocities between times with and without a current sheet for the simulations that allow for such a comparison (ignoring simulations with only initial, transient current sheets). We observe a general trend of stronger mid-plane turbulence being accompanied by a stronger magnetic field (smaller $|\beta_0|$). At times when the toroidal field exhibits a current sheet, the mid-plane turbulence strength is a factor of $\sim$1.5--3 stronger than in times without a current sheet. One exception is the simulation with $R = 30$ AU, $\beta_0 = -10^4$, where the mid-plane turbulence is $\sim$5--20\% larger at times without a current sheet.

Previous shearing box studies centered in the inner disk \citep{Bai2013_wind1,Lesur2014,Bai2015} employed a lower numerical resolution than this work, and thus have more numerical diffusion. It is possible that a larger numerical diffusion facilitates the ejection of current sheets on shorter timescales. To test this, we compared our standard resolution runs with those of a lower resolution (8 grid cells per $H$) and found that the lower resolution runs exhibit current sheets that are less complex and much shorter-lived, ensuring an even toroidal field geometry at later times. Whether or not a current sheet manifests in a shearing box simulations appears to be stochastic (\citealt{Bai2013_wind1}; see their \S4.4), and although we speculate (along with \citealt{Lesur2014}) that current sheets are ultimately transient on long enough timescales, we have seen that current sheets can be generated spontaneously (this work; \citealt{Bai2015}); thus the strength of turbulence in real disk systems may from time to time be enhanced by current sheets.

\subsection{Effects of Cosmic Ray Flux}
\label{sec:results:cosmic_rays}

As the cosmic ray flux through the disk is highly uncertain (see \S\ref{sec:method:ionization} above), we also examine the effect of a decreased cosmic ray ionization rate ($\zeta_{\rm CR,0} = 10^{-17}$; \citealt{Umebayashi1981}), as in e.g. \citet{Bai2011_ambipolar,Bai2015,Simon2018} (Figure \ref{fig:bfield_param_deltav}). Magnetically driven turbulence should be weaker with a lower degree of coupling between the gas and magnetic field. Despite this, we find that the turbulence is fairly insensitive to the degree of cosmic ray ionization -- this comparison is done only for times when a current sheet is not present, to control for the current sheet's influence on the turbulence. A possible explanation for this is that although cosmic rays are the dominant source of ionization at the mid-plane in our disk model, reducing the cosmic ray flux by a factor of 10 only reduces the ionization fraction by a less drastic factor $\sim 3$ under the metal-and-grain-free dissociative recombination approximation of \citet{Gammie1996} and \citet{Fromang2002}. The effects of weaker coupling of the magnetic field to the gas motions may only become apparent with a lower cosmic ray flux than considered here.

\subsection{Effects of Radial Disk Location}
\label{sec:results:radius_param}

\begin{figure}
    \centering
    \includegraphics[width=\columnwidth]{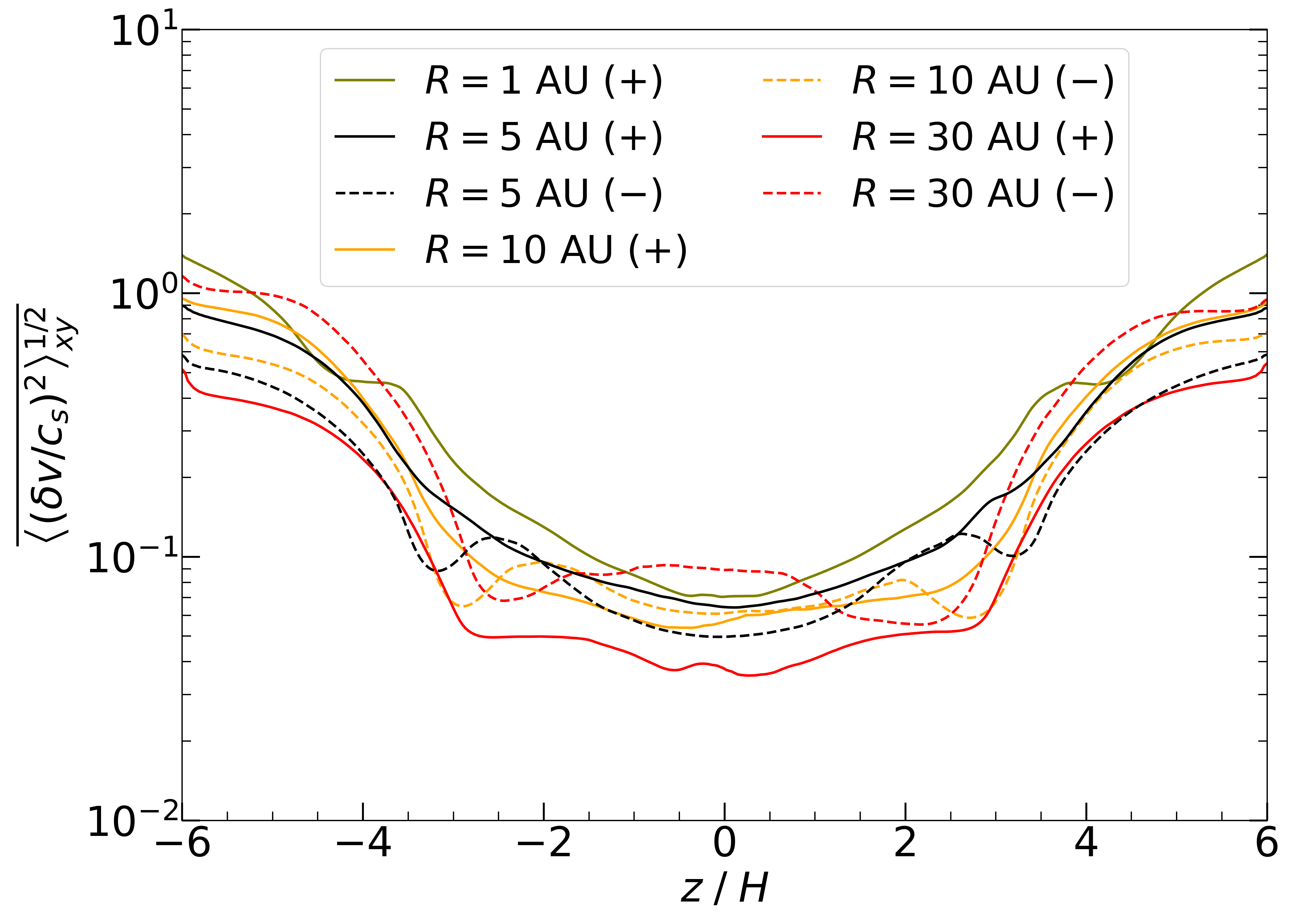}
    \caption{Same as Figure \ref{fig:velocity_profile}, but varying the disk radial location. Solid lines and $(+)$ denote $+\beta_0$, and dashed lines and $(-)$ mark $-\beta_0$. Between 1--30 AU, the gas turbulence appears relatively insensitive to radial location and initial magnetic field polarity.}
    \label{fig:radius_param_deltav}
\end{figure}

We investigate the effects of radial disk location on the magnetic field and velocity structure at $\beta_0 = \pm 10^4$ (Figure \ref{fig:radius_param_deltav}). Between 1--30 AU mid-plane values are between $\sim 0.04$--$0.1$ $c_s$ regardless of the polarity of the initial magnetic field, and without a clear trend across the radii considered. At 1 AU, the turbulent velocities exceed the sound speed at the disk surface.

%
%
\section{Discussion}
\label{sec:discussion}

\subsection{Origin of the Turbulence}
\label{sec:discussion:origin}

The strength of the initial vertical magnetic field $B_{z,0}$ and the presence (or lack of) a current sheet are the two factors with the largest effect on the strength of the gas turbulence in our simulations (Figures \ref{fig:bfield_param_deltav} and \ref{fig:cs_comparison}). This indicates that the turbulence in our simulations is primarily generated by magnetic effects. Between 1--30 AU (depending on the disk model), the Hall effect is the dominant non-ideal magnetic effect (Figure \ref{fig:elsasser_profiles}); when the Hall effect is strong, as it is in our simulations, the orientation of the magnetic field is also important \citep{Balbus2001,Kunz2008,Lesur2014,Bai2015,Simon2015_numeric}. The influence of the magnetic field polarity can be seen when comparing the $+$ and $-\beta_0$ configurations above at 5 AU (\S\ref{sec:results:magnetic_field_structure}), as the toroidal field and Maxwell stress are qualitatively and quantitatively distinct between the two cases. This is evidence that the Hall effect has a strong presence in our simulations. This difference is also apparent at 10 AU, but much less pronounced at 30 AU, where the Hall effect is weaker; therefore it may be difficult to obtain observational signals of Hall MHD at larger radii.

Previous works (e.g., \citealt{Lesur2014}) have shown that the HSI \citep{Kunz2008} produces strong, laminar toroidal magnetic fields (when $\beta_0 > 0$), which we also observe in the current set of simulations. In addition to this, our simulations exhibit large scale, laminar azimuthal velocities (Figure \ref{fig:midplane_velocities}) which are strong \emph{only} when $\beta_0 > 0$. This could indicate that the HSI is the origin of these large scale flows (or that an anti-aligned magnetic field prevents the magnetic self-organization like that seen by \citealt{Riols2019} for Ohmic and ambipolar diffusion only). Despite this, there is a strong turbulent gas component that is insensitive to the initial magnetic field orientation. In addition to the large scale, axisymmetric HSI, the presence of which is determined only by the sign of $\beta_0$, other ``guises'' of the HSI may be at work.

The non-axisymmetric HSI (hereafter, naHSI) \citep{Kunz2008,Simon2015_numeric} occurs when whistler waves become physically decoupled from the epicyclic motion of radially perturbed bulk neutral fluid, and may be driven unstable if magnetic perturbations grow by Keplerian shear faster than they can be advected by \alfven waves. The dispersion relation that described this behavior can be found in \citet{Simon2015_numeric}. Eventually stabilizing terms dominate over destabilizing influences; in other words, the naHSI is inherently a transient instability analogous to the non-axisymmetric MRI \citep{Balbus1992}.
\citet{Simon2015_numeric} found that the naHSI was likely responsible for sporadic bursts of magnetic stress in their simulations, even while $(\vec{\Omega}\cdot\vec{B}) < 0$, which lends credence to the possibility that this version of the HSI may be present in our simulations regardless of magnetic field polarity.

In addition to the naHSI, a local version of the axisymmetric HSI may generally survive (on scales less than the domain size) even when $\beta_0 < 0$. Although differential Keplerian rotation $\vec{\Omega}$ is the single largest source of shear in the disk, weaker shear exists between any adjacent fluid parcels with different vorticity. Similarly, the magnetic field will evolve beyond the initial vertical field to have an arbitrary orientation in three dimensions (as in Figure \ref{fig:bfield_xz}). Therefore, the presence of the HSI in a local region may not be determined only by the sign of $\vec{\Omega}\cdot\vec{B}$. Thus, the HSI may locally persist at arbitrary orientations in our simulations regardless of the background magnetic field polarity.

Beyond the possibility of the naHSI or a local version of the HSI, it is not clear if one or more secondary mechanisms are at work generating turbulence in our simulations. We will more fully follow up on these possibilities and investigate the origin of the turbulence in a subsequent paper.

\subsection{Impact on Particle Growth and Planetesimal Formation}
\label{sec:discussion:planet_formation}

Although strong Ohmic diffusion like that found in the inner disk will suppress MHD turbulence driven by the MRI, the Hall effect may still generate turbulence, replacing the classical ``dead zone'' (as in \citealt{Gammie1996,Balbus2001,Kunz2004}). In this case, even if disk accretion is primarily driven by magneto-thermal winds, significant turbulence can persist near the disk mid-plane. The existence of this turbulence may have significant implications for the growth of dust grains in planet-forming disks. The mid-plane turbulence levels found in this work are well above the fragmentation limit of \citet{Guttler2010} for collisions between compact grains of equal sizes. More recently, \citet{Kothe2016} (see also a review by \citealt{Simon2022}) found that grain growth could not proceed between equally sized silicate or water ice aggregates for collisional speeds of 25--50 ${\rm m\;s^{-1}}$ (3.8--7.5$\times 10^{-2}~c_s$ at 5 AU in our disk model). The strong velocities produced in our simulations may impede the collisional growth of grains; however, collisional grain growth might still occur at these speeds between grains or aggregates of different sizes or porosity (e.g., mass transfer, sticking; see \citealt{Guttler2010} for details).

The levels of turbulence found in this work may greatly reduce the efficacy of planetesimal formation by preventing the clumping of particles via, e.g., the streaming instability either by affecting the linear modes of the instability \citep{Umurhan2020,Chen2020} or preventing clumping in the non-linear state \citep{Gole2020,Lim2023_arxiv}.
For instance, while \citet{Li2021} found the critical mid-plane particle-to-gas density threshold $\epsilon_{\rm crit}$ for planetesimal formation via the streaming instability to be $\sim$ 0.3--1 (depending on the particle size), \citet{Lim2023_arxiv} found that driven turbulence imposes a stricter threshold of $\epsilon_{\rm crit} \sim $ 1--3 for turbulent velocities as low as $0.01~c_s$ --- the turbulent velocities found in this work greatly exceed these levels. However, \citet{Yang2018} demonstrated that anisotropies in the turbulent flow can actually head to enhanced radial concentration of particles despite a lack of sedimentation toward the mid-plane.

\subsection{Uncertainties and Caveats}
\label{sec:discussion:uncertainties}

\subsubsection{Dust Grains}
\label{sec:discussion:uncertainties:dust_grains}

Our setup uses a generous ionization prescription. While we might expect some disks to have (relatively) high levels of ionization, perhaps e.g. during an FU-Orionis-type stellar outburst, the precise ionization levels at the disk mid-plane are highly uncertain. While our results show that the strength of turbulence is relatively insensitive to the precise level of cosmic ray flux, it is important to consider other ways in which the ionization can be impacted. Dust grains are the electron absorbers that have the single largest effect on ionization. Including a self-consistent treatment of dust grains or grain growth in our simulations is beyond the scope of this paper, but here we briefly consider the effect dust grains might have on our results.

Dust grains are a catalyst for recombination \citep{Weisheit1978} and ``soak up'' electrons and ions from the gas, reducing the gas conductivity by several orders of magnitude \citep{Wardle1999,Wardle2007,Bai2011_MRI,Bai2011_tinygrains} and by extension altering the strength of low-ionization MHD effects. In particular, the Hall effect will be amplified by reduced gas conductivity (Equation \ref{eq:eta_Ha}): the drift velocity between electrons and ions (Hall current) will increase as the number density of neutrals, which induce a collisional drag on the ions, is increased (though this trend cannot continue indefinitely -- as the number of neutrals increase, the gas will increasingly decouple from the magnetic field and approach a purely hydrodynamic system). Grain-free treatments (e.g. \citealt{Gammie1996,Fromang2002}; this work) assume that dust has completely settled to the disk mid-plane and thus does not affect the dynamics or ionization of the bulk gas. However, in general, turbulence will diffuse dust grains away from the mid-plane \citep{Fromang2006} and mix grains into the bulk gas. In a follow up paper, we will further study the strength and nature of gas turbulence for weaker ionization levels.

\subsubsection{Numerical Limitations}
\label{sec:discussion:uncertainties:numerical_lims}

As with any study relying on local simulations, we must ask whether the observed behavior will appear in a global setup. While shearing boxes are unsuited for studies of large scale structures (e.g., winds), turbulence is inherently a small-scale, local phenomenon. Likewise, global simulations that strive to capture both large and small-scale phenomena are often too computationally intensive. Early global simulations that included non-ideal MHD suggested that between $\sim$1--10 AU, the disk is laminar \citep[e.g.,][]{Gressel2015,Bethune2017} with accretion driven primarily by magnetically launched winds. However, these models employ much lower resolution (respectively, 24 and 16 grid cells per scale height at their most refined) and both use 2D (axisymmetric) domains (although \citet{Bethune2017} do consider a few 3D simulations). Fundamental differences between 2D and 3D turbulence and low resolution may have affected the results of these early global simulations. More recent 3D, non-ideal MHD global simulations focusing on the outer disk ($> 30$ AU) have employed static mesh refinement \citep{Bai2017} to better resolve small-scale turbulence (if present) in the bulk of the disk.  These simulations produced turbulent velocities that are in good agreement with shearing box simulations \citep{Simon2018} centered in the outer disk and with comparable resolution. Thus, it is possible that the earlier global simulations show little turbulence due to the lower resolution (e.g., no static mesh refinement) not adequately resolving turbulent velocity fluctuations at small scales.

One limitation of our setup is the radial symmetry of the shearing box (see \citealt{Bai2013_wind1,Lesur2014}), which can result in vertical outflows that are radially bent in opposite directions at the top and bottom vertical boundary (see Figure \ref{fig:turbvel_xz}) when there is no current sheet in the toroidal magnetic field. Real disks and global simulations do not have radial symmetry, and thus should always have a current sheet (i.e. magnetic fields and outflow are always bent away from the central star). Whether or not a current sheet develops in a shearing box is highly stochastic \citep{Bai2013_wind1}. In a global simulation, this radial symmetry is broken, thus the toroidal magnetic field should always exhibit an overall sign change across a current sheet in the vertical direction, with poloidal field lines bent away from the central protostar (though this current sheet will not necessarily be at the mid-plane either, e.g. \citealt{Bai2017,Hu2023}). Because of the local box approximation of the simulations in this work, it is possible for current sheets to escape the simulations domain.

Lastly, our shearing box setup is isothermal. Such a setup cannot capture the turbulence generated by effects that rely on temperature gradients or gas cooling (e.g. \citealt{Lyra2019}). However, our setup is particularly advantageous for identifying turbulence generated exclusively by non-ideal MHD.

\subsubsection{Consistency with Disk Observations}
\label{sec:discussion:uncertainties:observations}

There is a shortage of observations of turbulence in the 1--10 AU regions of protoplanetary disks with which to compare our simulations. The most direct constraints on turbulence come from the turbulent broadening of molecular emission lines (e.g., \citealt{Flaherty2017,Flaherty2018,Flaherty2020}) but these studies are limited by spatial resolution to large disk radii $\gtrsim 30$ AU, and in most cases limited to the uppermost surface layer of the disk by the large optical depth of the molecular line being observed (e.g., CO 3--2 rotational line). While spatial resolutions of 5 AU have been achieved for mm continuum emission (e.g., DSHARP; \citealt{Andrews2018,Franceschi2023,Pizzati2023}) to our knowledge, these observations have not been used to constrain turbulence directly via line-broadening.
Continuum observations of dust ring width can be used to constrain turbulence (e.g., \citealt{Rosotti2020,Jennings2022}) but, these attempts must still resolve structures on the order of the gas scale height, i.e. $\sim 4$~AU at a radius of 40 AU, which means that turbulence can only be constrained in this way for radii $\gtrsim 40$ AU. The MAPS observations resolved emission lines at $\sim 20$ AU scales (MAPS; \citealt{Oberg2021,Sierra2021}), thus still limiting constraints to the outer disk. Finally, \citet{Carr2004} used hot rovibrational H$_2$O emission to find strong turbulence near the disk surface, but at a small radius of 0.3 AU. Indirect measurements of turbulence via modelling the spectral energy distributions of young stellar objects (e.g. \citealt{Mulders2012}) are thus far unable to distinguish between strong and weak turbulence, given degeneracies between the strength of turbulence and other parameters, such as the dust-to-gas ratio.

The simulations presented here demonstrate mid-plane turbulent velocities between $\sim3$--$9\times 10^{-2}c_s$ across all tested values of $\beta_0$ and radii.
Line broadening observations of optically thick lines (thus constraining these measurements to well away from the disk mid-plane) place upper limits of the turbulence in the outer disk of HD 163296 to be $\lesssim 0.05c_s$ (\citealt{Flaherty2017}; in this work, turbulent gas motions near the mid-plane were also quantifiable) and TW Hydra to be $\lesssim 0.08c_s$ \citep{Flaherty2018}. \citet{Flaherty2020} find strong turbulence (0.25--0.33$c_s$) around DM Tau, consistent with earlier work by \cite{Guilloteau2012}, who found a range of turbulent velocities 0.12--0.69$c_s$. Furthermore, \cite{Flaherty2020} place upper limits of $< 0.08c_s$ and $< 0.12c_s$ around MWC 480 and V4046 Sgr, respectively. More recently, \cite{PC2023} found turbulent velocities of $\approx 0.4$--$0.6c_s$ at relatively large heights above the mid-plane ($z/r = 0.2$--0.3, where $z (r)$ is the vertical (radial) coordinate of the disk) in IM Lup.
The level of turbulence observed in the outer disk surface (where ambipolar diffusion is the dominant low ionization effect) cannot be compared in a precise manner to the turbulence level found in this work (in the inner disk, where the Hall effect is dominant). However, the turbulence limits constrained by such observations (e.g., $\lesssim 0.03c_s$, $\lesssim 0.08 c_s$) are much smaller than the turbulent velocities produced in our simulations near the disk surface, which approach sonic speed. Thus, if such turbulent velocities are present in physical disks, they are only present in the inner disk where the Hall effect is much stronger than ambipolar diffusion. Observations with spatial resolutions that can probe the innermost radial regions ($R \lesssim 5$ AU) of disks are needed to verify that such turbulence is present.

%
%
\section{Summary}
\label{sec:conclusions}

We performed 3D local shearing box simulations representing the inner 1--30 AU of a model protoplanetary disk including all three non-ideal MHD effects (Ohmic diffusion, the Hall effect, and ambipolar diffusion). These simulations have the highest resolution per gas scale height among similar simulations to date, with 32 zones per $H$ in every dimension. We explored how turbulence is affected by magnetic strength and polarity as well as radial location within the disk. Our key results are as follows:

\begin{itemize}
    \item In the presence of the Hall effect, weakly ionized disks can produce vigorous gas turbulence, with turbulent velocity fluctuations of at least $\sim 3$--$9\times 10^{-2} c_s$ (16--75 ${\rm m\;s^{-1}}$ in our disk model) at the mid-plane and $\sim c_s$ (0.4--1 ${\rm km\;s^{-1}}$) near the disk surface. 
    \item This turbulence is sensitive to the initial magnetic field strength and the presence or absence of a current sheet associated with a flip in the horizontally averaged toroidal field. As such, the turbulence is largely magnetically-driven in nature.
    \item Despite the role of magnetic fields in generating turbulence, the strength of the turbulence is not sensitive to the polarity of the initial vertical field.
    \item The magnetic field is dominated by a strong, laminar toroidal component, with a weaker, turbulent poloidal component. Often, but not always, a current sheet is present near the mid-plane.
    \item The mid-plane Maxwell stress is extremely sensitive to the polarity of the initial vertical magnetic field.
\end{itemize}

Overall, these findings show that gas in the inner (e.g, 1--30 AU) regions of protoplanetary disks can be strongly turbulent. As mentioned previously, there are several caveats to bear in mind. The neglect of dust grains and the physically motivated but generous ionization prescription present in this work may not reflect the conditions present in real protoplanetary disks. In addition, the toroidal field current sheet, which we have shown can cause drastic variability in the levels of turbulence, may not be adequately modeled without both a global domain and an even higher resolution. While these are clear uncertainties to address in future works, our results demonstrate the importance of considering the detailed gas dynamics in modeling such disks even if the accretion flow is mostly laminar and dominated by magnetically driven winds.

%
%

\section*{Acknowledgments}

We thank Kevin Flaherty, Meredith Hughes, and Xiao Hu for useful discussions related to this work.
We acknowledge support from NASA via the Theoretical and Computational Astrophysics Network (TCAN) program through grant 80NSSC21K0497.
D.G.R. and J.B.S acknowledge support from NASA under the Emerging Worlds program through grant 80NSSC20K0702.
D.C. acknowledges support from NASA under the Emerging Worlds program through grant 80NSSC21K0037.
G.L. acknowledges support from the European Research Council (ERC) under the European Union's Horizon 2020 research and innovation program (Grant Agreement No. 815559 (MHDiscs)).
W.L. acknowledges support from the NASA Emerging Worlds program via grant 80NSSC22K1419 and by NSF via grant AST-2007422.
C.C.Y. is grateful for the support from NASA via the Astrophysics Theory Program (grant \#80NSSC21K0141) and the Emerging Worlds program (grant \#80NSSC20K0347 and \#80NSSC23K0653).
The computations were performed using Stampede2 at the Texas Advanced Computing Center using XSEDE/ACCESS grant TG-AST120062.

\software{Athena \citep{Stone2008}, NumPy \citep{harris2020array}, Matplotlib \citep{Hunter:2007}}

%
%

\bibliography{references}{}
\bibliographystyle{aasjournal}

\end{document}